\documentclass[12pt]{article}

\pdfoutput=1

\usepackage{amssymb,amsfonts}
\usepackage[all,arc]{xy}
\usepackage{enumerate}
\usepackage{mathrsfs}
\usepackage{mathtools}
\usepackage{amsmath}
\usepackage{lipsum}
\usepackage{changepage}
\usepackage{upgreek}
\usepackage{bm}
\usepackage{chngcntr}
\usepackage{graphicx}

\usepackage{pifont}
\newcommand{\cmark}{\ding{51}}
\newcommand{\xmark}{\ding{55}}

\usepackage{hyperref}
\usepackage{cleveref}
\hypersetup{colorlinks, citecolor=blue}

\usepackage{apacite}
\makeatletter
\newfont{\footsc}{cmcsc10 at 8truept}
\newfont{\footbf}{cmbx10 at 8truept}
\newfont{\footrm}{cmr10 at 10truept}

\renewcommand{\ps@plain}{%
\renewcommand{\@oddfoot}{\footsc
 (2024), Dasylva, Goussanou and Nambeu  \hfil\footrm\thepage}}
\makeatother

\title{Models of linkage error for capture-recapture estimation without clerical reviews}

\author{A. Dasylva\footnote{abel.dasylva@statcan.gc.ca}, A. Goussanou and C.-O. Nambeu\\ Statistics Canada}

\date{\small Written: Mar. 2024}

\begin{document}

\maketitle

\newtheorem{theorem}{Theorem}
\newtheorem{lemma}{Lemma}
\newtheorem{corollary}{Corollary}


\begin{abstract}
The capture-recapture method can be applied to measure the coverage of administrative and big data sources, in official statistics.
 In its basic form, it involves the linkage of two sources while assuming a perfect linkage and other standard assumptions.
 In practice, linkage errors arise and are a potential source of bias, where the linkage is based on quasi-identifiers.
 These errors include false positives and false negatives, where the former arise when linking a pair of records from different units, and the latter arise when not linking a pair of records from the same unit.
 So far, the existing solutions have resorted to costly clerical reviews, or they have made the restrictive conditional independence assumption.
 In this work, these requirements are relaxed by modeling the number of links from a record instead.
 The same approach may be taken to estimate the linkage accuracy without clerical reviews, when linking two sources that each have some undercoverage.
\end{abstract}


\vspace{6pt}

{\footnotesize
\noindent {\bf Disclaimer}: The content of this paper represents the authors' opinions and not necessarily those of Statistics Canada.}



\section{Introduction}

The capture-recapture method is an important tool for estimating the coverage of administrative and big data sources that are increasingly used in official statistics \cite{lichun_zhang_2015}.
 In its simplest form, it estimates the coverage of two sources on the same finite population, by identifying the units selected in both sources, i.e., their intersection, under standard assumptions that include a perfect linkage.
 Then, the estimated coverage is based on the well known estimator by \citeA{petersen_1896} and \citeA{lincoln_1930}.
 However, linkage errors may arise because the linkage is often based on quasi-identifiers such as names and dates.
 These errors may bias the coverage estimate, which must be corrected.


Regarding this accuracy, a linkage error is defined as a {\it false negative} or a {\it false positive}, where a false negative is failing to link records from the same unit, and a false positive is linking records from different units.
 In connection with these concepts, a record pair is called {\it matched} if its records are from the same unit \cite{fellegi_sunter_1969, herzog_scheuren_winkler_2007}.
 Otherwise, it is called {\it unmatched}.
 The linkage accuracy may be measured by clerical review, a statistical model, or a combination of both approaches.
 Clerical reviews consist in the visual inspection of a probability sample of record pairs to determine if they are matched \cite{dasylva_et_al_2016}.
 They are very flexible and apply regardless of the linkage details.
 However they are costly.
 The alternative to clerical reviews is fitting a statistical model of which quite a few have been proposed, including log-linear mixtures \cite{fellegi_sunter_1969,thibaudeau_1993,winkler_1993,daggy_et_al_2013,chipperfield_et_al_2018,winglee_et_al_2005,chipperfield_chambers_2015,haque_et_al_2021,haque_mengersen_2022}, models of a pair probabilistic linkage weight \cite{belin_rubin_1995,sariyar_et_al_2011}, Bayesian models \cite{fortini_et_al_2001,tancredi_liseo_2011,sadinle_2017,steortS_et_al_2016}, and models based on the number of links from a given record \cite{blakely_salmond_2002,dasylva_goussanou_jjsds_2022}.
 This latter modeling approach is of special interest in this work, because it is not limited to probabilistic linkages and implicitly accounts for all the interactions among the linkage variables.
 The modeling approach is not as costly as clerical reviews, but it is less flexible as it relies on assumptions about the linkage procedure.
 It is also quite challenging when the linkage is constrained to have at most one or exactly one link per record \cite[p. 226]{lahiri_larsen_2005}.
 \citeA{chipperfield_chambers_2015}, and \citeA{sadinle_2017} have addressed this issue.
 However, the proposed methodologies are computer intensive and depend on the restrictive assumption that the linkage variables are conditionally independent, i.e., they are independent given that a pair is matched or unmatched.
 Indeed, this assumption is a potential source of bias, according to \citeA[chap E.6, p. 149]{newcombe_1988}, \citeA{belin_rubin_1995} and \citeA{blakely_salmond_2002}.
 Following \citeA{larsen_rubin_2001}, it is also possible to combine clerical reviews and statistical modeling to take advantage of the flexibility of the former and the low costs of the latter.
 However, the overall costs remain beyond the budget of many studies.
 From the point of view of the capture-recapture method, linkage errors are detrimental because they may bias the estimated coverage.
 Indeed, a false negative may lead to underestimating the coverage, while a false positive may produce a bias in the opposite direction.
 Of course, this bias must be removed to accurately estimate the coverage.


Many error correction methods have been described, which make the standard capture-recapture assumptions except for the imperfect linkage, i.e., a closed population, independent units that are selected independently by each source, a homogeneous capture probability by at least one source, and no duplicates or out-of-scope units in either source.
 They include solutions that require clerical estimates of the linkage accuracy \cite{ding_fienberg_1994,diconsiglio_tuoto_2015,dewolf_et_al_2019,brown_et_al_2020}, and other solutions that rely on a statistical model under the conditional independence assumption \cite{tancredi_liseo_2011,racinskij_et_al_2019}.
 \citeA{ding_fienberg_1994}, \citeA{diconsiglio_tuoto_2015} and \citeA{dewolf_et_al_2019} describe three closely related solutions of the former kind, where they constrain the linkage to have at most one link per record, and assume that the false positive probability is negligible for units that are captured by both sources.
 However, they estimate the linkage accuracy through clerical reviews,  which are costly but the only practical solution, given the linkage constraints.
 \citeA{brown_et_al_2020} describe a different solution, which also relies on clerical estimates of the linkage accuracy, where the two sources must be linked twice with different linkage procedures, under the assumption that the related link indicators are independent in each matched pair.
 Instead, \citeA{tancredi_liseo_2011}, and \citeA{racinskij_et_al_2019} use statistical models to jointly estimate the linkage accuracy and the coverage without clerical reviews.
 However they make the restrictive conditional independence assumption.


This work aims to jointly estimate the coverage and linkage accuracy without clerical reviews, while relaxing the assumption that the linkage variables are conditionally independent.
 To that end, a new methodology is described, which extends a previous model of linkage error \cite{dasylva_goussanou_jjsds_2022}, under the standard capture-recapture assumptions except for the perfect linkage assumption. 
 In this model-based approach, the coverage is estimated by linking the records with a sufficiently high recall, or by specifying the interactions in the matched pairs, while allowing arbitrary interactions in the unmatched pairs.
 The same models may be used to estimate the recall and precision when linking two sources that each have some undercoverage.

 The remaining sections comprise the notations and assumptions, background, proposed methodology, simulations, and conclusion, in this order.


\section{Notations and assumptions}

In the basic version of the capture-recapture problem, the coverage of a list from a finite population must be estimated by exploiting a second list from the same finite population, under standard assumptions, which include a closed population, independent units that are selected independently by each list, homogeneous capture by at least one list, no duplicates or out-of-scope units in either list and a perfect linkage of the two lists.
 In general, these lists correspond to probability samples with unknown selection probabilities.
 In what follows, it is assumed that the standard capture-recapture assumptions hold except for the perfect linkage assumption that is relaxed.
 Each list is modeled by a Bernoulli sample, where each included unit is associated with a record that possibly contains typos.
 For example, with person lists, this record may contain the last name and the birth date.
 The record values are assumed to be independent across the units, but no assumption is made regarding the dependence of these values for records that are from the same unit, or the dependence of the variables within these records.
 To satisfy the homogeneous capture assumption, it is assumed that a unit is captured in the first list independently of the associated record values (e.g., the recorded last name and birth date).
 However, capture in the second list may depend on these values.
 In what follows, denote the cardinality of a set $s$ by $|s|$, and for a tuple $\bm{x} = \left ( x_1,\ldots,x_d\right ) \in \mathbb{R}^d$ let $|\bm{x}| = |x_1|+\ldots+|x_d|$.

\vspace{6pt}


\subsection{Finite population and data sources}\label{subsection: finite population and sources}

The units are from a finite population $U$ with $N$ units that are labeled from $1$ to $N$.
 The units are selected according to two Bernoulli samples that are denoted by $S_A$ and $S_B$ and identified by two subsets of $\{1,\ldots,N\}$, where it is assumed that the inclusion probability $P(i \in S_A)$ does not depend on $N$.
 For example, $S_A$ may be the census of population and $S_B$ may be a coverage survey.
 In each list where it is included, unit $i$ is associated with a record, whose value is its defining characteristic.
 In what follows, we let the term record also refer to this value where it is clear from the context.
 The record value is assumed to live in the record space ${\cal V}_N$, which is finite but possibly large.
 For example, the record space may comprise all the strings written with no more than 32 alphabetical characters, if linking with  the surname.
 In $S_B$, this record is denoted by $V_i$.
 For $S_A$, the labeling of the records depends on a uniformly random permutation $\Pi(.)$ of $\{1,\ldots,N\}$, to model the complete lack of information about the records associated with the same unit.
 In this list, unit $i$ is associated with the record $V_{\Pi(i)}'$.
 The use of a random unknown permutation is a common device in the record linkage literature \cite{lahiri_larsen_2005,chambers_2009a}.
 It is mathematically convenient to view the two lists as samples drawn from conceptual registers $A$ and $B$, where each unit is associated with a record.
 Then, the recording and list inclusion processes are assumed to be such that the observations
%
%
\[ \left [ \left ( I(i \in S_A ),I(i \in S_B ),V_i,V_{\Pi(i)}' \right ) \right ]_{1 \leq i \leq N} \]

\noindent are independent and identically distributed and independent of the random permutation $\Pi(.)$.
 This means that the two lists are labeled independently and that one may consider only the case where the permutation $\Pi(.)$ is the identity (i.e., conditioning on $\Pi(.)$ being the identity in what follows), without loss of generality.
 Then, unit $i$ is associated with $V_i$ in $S_B$ and $V_i'$ in $S_A$.
 In the discussion that is to follow, all the arguments are conditional on $\Pi(.)$ being equal to the identity.
 However, this information is omitted to simplify the notation.
 To satisfy the homogeneous capture assumption, it is assumed that capture in $S_A$ is independent of $V_i$ and $V_i'$.
 However capture in $S_B$ may depend on these records, i.e. we may have
%
%
\[ P \left ( i \in S_B \left | V_i,V_i'\right . \right ) \neq
   P(i \in S_B ). \]
 For example, the capture probability may vary across post-strata based on $V_i$.

\vspace{6pt}



\subsection{Record linkage and related errors}

The indicator of a link between $V_i$ and $V_j'$ is denoted by $L_{ij}$ and called {\it linkage decision} for the pair $(i,j)$.
 Let $n_i$ denote the number of links from $V_i$ in $S_B$, i.e.,
%
%
\begin{equation}
 \label{eq.: general n_i}
 n_i = \sum_{j \in S_A} L_{ij}.
\end{equation}
 The linkage is assumed to be such that $L_{ij}$ is only a function of $V_i$ and $V_j'$, i.e., the decision to link two records does not involve other records.
 In the current setup, this assumption precisely means that
%
%
\begin{equation*}
 E \left [ L_{ij} \left |
 \begin{array}{l}
 \left [ \left ( I(k \in S_A ), I(k \in S_B ), V_k,V_k' \right )
 \right ]_{k \in \{1,\ldots,N\}\setminus \{i,j\}}, \\
 (i,j) \in S_A \times S_B, V_i,V_j'
 \end{array} \right . \right ] =
 E \left [ L_{ij} \left | (i,j) \in S_A \times S_B,
 V_i,V_j'  \right . \right ].
\end{equation*}
 This condition covers a wide range of practical linkage strategies that may be implemented with the probabilistic, deterministic or hierarchical methods.
 However it excludes linkages that constrain the number of links per record (e.g., exactly one or at most one) even if such linkages may be built from simpler ones, which meet the condition.
 A record pair is denoted by an element $(i,j)$ of $S_A \times S_B$.
 As mentioned before, a pair is called {\it matched} if the two records are from the same unit.
 Otherwise, it is called {\it unmatched}.
%
%
%
 To discuss the linkage errors, a {\it false negative} is a matched pair that is not linked, a {\it false positive} is an unmatched pair that is linked, and a {\it true positive} is a matched pair that is linked.
 For completeness, define a {\it true negative} as an unmatched pair that is not linked.
 For convenience, let $FN, FP$, $TP$ and $TN$ denote the total numbers of false negatives, false positives, true positives and true negatives, respectively.
 It is common to represent the different kinds of record pairs in a $2\times 2$ table called a confusion matrix where the off-diagonal cells represent the errors, as shown in Table~\ref{table: confusion matrix}.
%
%
\begin{table}[hbt!]
\centering
\caption{Confusion matrix. \label{table: confusion matrix}}
\vspace{6pt}
{\small
\begin{tabular}{l|c|c|}\cline{2-3}
			  & Link & No link \\ \hline
\multicolumn{1}{|l|}{Matched}			  & $TP$	 & $FN$ \\ \hline
\multicolumn{1}{|l|}{Unmatched}			  & $FP$    & $TN$ \\ \hline
\end{tabular}}
\end{table}
 The linkage accuracy is typically measured by the recall and the precision, where the {\it recall} is the proportion of matched pairs that are linked (i.e. $TP/(TP+FN)$) and the {\it precision} is the proportion of linked pairs that are matched (i.e. $TP/(TP+FP)$).
 It is also measured by the {\it false negative rate}, which is the proportion of matched pairs that are not linked (i.e., $FN/(TP+FN)$), and the {\it false positive rate} (FPR), which is the proportion of unmatched  pairs that are linked (i.e., $FP/(TN+FP)$).
 With a perfect linkage, the precision and recall are equal to $1.0$, while the FPR is null.
 In this ideal situation, the population size is estimated according to \citeA{petersen_1896} and \citeA{lincoln_1930} by \[ \widehat{N} = \frac{|S_A| |S_B|}{|S_A \cap S_B|}.\]
 Consequently, the estimated coverage of $S_A$ is given by $|S_A \cap S_B|/|S_B|$, while that of $S_B$ is given by $|S_A \cap S_B|/|S_A|$.
 With linkage errors, the intersection of the two lists is not directly observed.
 Instead, the size of this intersection must be inferred from the observed links and the linkage accuracy that can be estimated by modeling the number of links from a given record.


\section{Background}

This section provides some background on the error model \cite{dasylva_goussanou_jjsds_2022}, which is to be adapted for the problem at hand.
 This model applies when $S_A$ is a census (i.e., $S_A=U$) and the linkage is such that the decision to link two given records involves no other record.
 In this case the linkage accuracy may be estimated by modeling the distribution of the number of links from a given record, without assumptions about the dependence of the linkage variables.

\vspace{6pt}



\subsection{How the errors relate to the number of links from a record}

In general, there is a strong connection between the number of links from a given record and the related linkage errors.
 This connection is described in Table~\ref{table: connection between n_i and the linkage errors for a census} when $S_A$ is a census \cite{dasylva_goussanou_2020}.
%
%
\begin{table}[hbt!]
\centering
\caption{Connection between $n_i$ and errors when $S_A$ is a census. \label{table: connection between n_i and the linkage errors for a census}}
\vspace{6pt}
{\small
\begin{tabular}{ccc}\hline
$n_i$			  & False negatives & False positives \\ \hline
0			  & 1	    & 0 \\
$1 \leq n_i \leq N-1$ & 0 or 1	    & $n_i-1$ or $n_i$ \\
$N$ & 0 & $N-1$ \\ \hline
\end{tabular}}
\end{table}
 When $n_i=0$, it is known that there is no false positive but one false negative because $S_A$ is a census.
 When $n_i=1,\ldots,N-1$, there is either no or one false negative and thus $n_i$ or $n_i-1$ false positives, according to whether the record is linked to the matched census record or not, because $S_A$ is a census  and it has no duplicate records.
 When $n_i=N$, it is known that there are no false negatives and $N-1$ false positives, for the same reasons.
 As an illustration, consider the example shown in Table~\ref{table: first example}, where $N=5$, $S_A=U=\{1,2,3,4,5\}$, $S_B=\{2,3\}$ and the nature of each record pair is also shown.
 In this example there are four links including $(2,1)$, $(2,2)$, $(3,2)$ and $(3,4)$.
 It can be verified that $n_i$ is related to the linkage errors according to Table~\ref{table: connection between n_i and the linkage errors for a census}.
 When $n_i=1,\ldots,N-1$, the errors are not fully known and may be predicted with a model.
%
%
\begin{table}[hbt!]
\centering
\caption{Example, where $N=5$, $S_A=U=\{1,2,3,4,5\}$ and $S_B=\{2,3\}$ and the links are indicated by the check marks.\label{table: first example}}
\vspace{6pt}
{\small
\begin{tabular}{rccccccccc}\cline{3-7}
	&		&	\multicolumn{5}{c}{$j=$}									&		&		&		\\ \cline{8-10}
	&		&	1	&	2	&	3	&	4	&	5	&	$n_i$	&	\# FN	&	\# FP	\\ \hline
$i=$	&	2	&	\cmark FP	&	\cmark TP	&	 TN	&	 TN	&	TN	&	2	&	0	&	1	\\
	&	3	&	 TN	&	\cmark FP	&	 FN	&	\cmark FP	&	TN	&	2	&	1	&	2	\\ \hline
\end{tabular}}
\end{table}

\vspace{6pt}



\subsection{A model for homogeneous records}

\citeA{blakely_salmond_2002} model $n_i$ by the sum of a Bernoulli variable (for the true positives) with an independent binomial variable (for the false positives) and they estimate the related parameters through a quadratic equation.
 However, the $n_i$ distribution must be the same for all the records.
 Otherwise, the estimator may be biased or fail to exist \cite{dasylva_goussanou_jjsds_2022} if the quadratic equation has no solution.
 In practice, this issue may arise when linking with names or other characteristics, which occur with different frequencies in the population.

\vspace{6pt}


\subsection{A model for heterogeneous records}\label{subsection: model for heterogeneous records}

To address the problem, \citeA{dasylva_goussanou_jjsds_2022} have extended the model from \cite{blakely_salmond_2002} into a finite mixture, which applies when $N$ gets large under regularity conditions.
 To describe these conditions, let
%
%
\begin{equation}
 {\cal V}_N^{\ast}= \left \{ v \in {\cal V}_N \ s.t. \ 
 P \left ( V_i = v \left | i \in S_B \right . \right )>0 \right \} .
\end{equation}
 In other words, ${\cal V}_N^{\ast}$ is the subset of record values that may be observed in $S_B$, with a positive probability.
 At this point, it is useful to consider the subset of all record values (from ${\cal V}_N$) that are linked to a particular record value with a positive probability, as well as a superset of this set, which is called {\it neighborhood} and denoted by ${\cal B}_N(v)$ for the value $v \in {\cal V}_N^{\ast}$.
 Thus,
%
%
\begin{equation}
 {\cal B}_N(v) \supset
 \left \{ v' \in {\cal V}_N \ s.t. \
 E \left [ L_{ij} \left |
 i \in S_B, \left ( V_i,V_j'\right )=(v,v') \right . \right ]>0
 \right \}.
\end{equation}
 Informally, the neighborhood of a particular record value is a subset of record values that look like this value according to some criterion.
 For example, consider linking records based on the last name in capital letters and suppose that two records are linked if they agree exactly on this variable.
 In this case the record value ($v$) ``JARO'' may be associated with the singleton neighborhood (${\cal B}_N(v) = \{v\}$) $\{\mbox{``JARO''} \}$.
 To refine this example, suppose now that two records are linked if the last names are identical, or they have the same length and differ by a single letter.
 In this case, the value ``JARO'' may be associated with the neighborhood
%
%
\begin{equation*}
\begin{array}{c}
 \left \{ \mbox{``AARO''}, \mbox{``BARO''}, \ldots,
          \mbox{``ZARO''}\right \} \cup
 \left \{ \mbox{``JARO''}, \mbox{``JBRO''}, \ldots,
          \mbox{``JZRO''}\right \} \cup \\
 \left \{ \mbox{``JAAO''}, \mbox{``JABO''}, \ldots,
          \mbox{``JAZO''}\right \} \cup
 \left \{ \mbox{``JARA''}, \mbox{``JARB''}, \ldots,
          \mbox{``JARZ''}\right \}.
\end{array}
\end{equation*}
 The concept of neighborhood is useful when characterizing the discriminating power of the linkage variables, and when articulating regularity conditions for the consistent estimation of the recall and precision without clerical reviews.
 To describe these conditions, define the functions $p_N(.)$, $\lambda_N(.)$ and $\lambda_N^{(0)}(.)$, which give the true positive probability, the false positive probability and the probability that an unmatched record is in the neighborhood, i.e.
%
%
\begin{eqnarray}
 p_N(v) &=&
 E \left [ L_{ii} \left | i \in S_B, V_i=v \right . \right ], \\
 \lambda_N(v) &=&
 E \left [ L_{ij} \left | i \in S_B, V_i=v \right . \right ],
 \ j \neq i , \\
 \lambda_N^{(0)}(v) &=&
 P \left ( V_j' \in {\cal B}_N \left ( V_i \right ) \left |
 i \in S_B, V_i=v \right . \right ).
\end{eqnarray}
 Then the first two regularity conditions are given by the following equations, where $\Lambda$ is positive and finite, $F$ is a bivariate distribution  with support contained in $[0,1] \times [0,\Lambda]$ and neither depends on $N$.
%
%
\begin{eqnarray}
 \label{eq: census regularity condition 1}
 \sup_{v \in {\cal V}_N^{\ast}} (N-1) \lambda_N^{(0)}(v) &\leq&
 \Lambda,\\
 \label{eq: census regularity condition 2}
 \left . \left ( p_N(V_i),(N-1)\lambda_N(V_i)\right ) \right |
 \{i \in S_B \} &\stackrel{d}{\longrightarrow}& F.
\end{eqnarray}
 The first condition implies that the expected number of false positives is bounded above for each record.
 When the true positive probability is bounded below by $\delta$ (Eq.~6 in \citeA{dasylva_goussanou_2020}), it also implies that the precision is no less than $\delta/(\delta+\Lambda)$ overall and for any post-stratum, which is defined based on $V_i$ \cite{dasylva_goussanou_2020}.
 The second condition means that the joint distribution of the expected number of true positives and the expected number of false positives is approximately given by $F$ when $N$ is large.
 When $F$ is discrete with $G$ atoms, these two conditions imply the following convergence in distribution \cite[Lemma 1]{dasylva_goussanou_jjsds_2022}.
%
%
\begin{eqnarray}
 \label{eq: limiting mixture}
 n_i \left | \{i \in S_B \} \right .
 &\stackrel{d}{\longrightarrow}&
 \sum_{g=1}^G \alpha_g Bernoulli (p_g) \ast Poisson (\lambda_g),
\end{eqnarray}
 where $\ast$ denotes the convolution operator. This means that, in the limit, a record belongs to one of $G$ latent classes, where $\alpha_g$ is the probability of class $g$, and $p_g$ and $\lambda_g$ are the expected numbers of true positives and false positives for the records in this class.
 The model parameters may be estimated by maximizing the composite likelihood of the $n_i$'s.
 They are related to the linkage accuracy through the expected numbers of true positives and false positives per record in $S_B$, which are given by $\overline{p} = \sum_{g=1}^G \alpha_g p_g$ and $\overline{\lambda} = \sum_{g=1}^G \alpha_g \lambda_g$.
 Indeed, the recall and precision converge in probability to ${\overline{p}}$ and ${\overline{p}}/({\overline{p}}+{\overline{\lambda}})$, respectively, under the following two additional regularity conditions \cite[Corollary 1]{dasylva_goussanou_jjsds_2022}, where $i \neq i'$ and $c$ is a positive finite constant not depending on $N$.
%
%
\begin{eqnarray}
 \label{eq: census regularity condition 3}
 N P \left ( \left . {\cal B}_N(V_i) \cap {\cal B}_N(V_{i'}) \neq
 \varnothing \right | \{i,i'\} \subset S_B \right ) \leq c,\\
%
%
 \label{eq: census regularity condition 4}
 N P \left ( \left . V_{i'}' \in {\cal B}_N(V_i) \right |
 \{i,i'\} \subset S_B, {\cal B}_N(V_i) \cap {\cal B}_N(V_{i'}) \neq
 \varnothing \right ) \leq c.
\end{eqnarray}
 These two conditions mean that records from different units are very likely to have disjoint neighborhoods (Eq.~\ref{eq: census regularity condition 3}) and that matched records are not far apart (Eq.~\ref{eq: census regularity condition 4}).
 With the other regularity conditions (i.e., Eqs.~\ref{eq: census regularity condition 1}-\ref{eq: census regularity condition 2}), they also imply the consistency of the composite maximum likelihood estimators \cite[Theorem 3]{dasylva_goussanou_jjsds_2022}.

Overall, this methodology has several advantages over alternative model-based solutions, because it seamlessly accounts for the interactions among the linkage variables and the records' heterogeneity.
 Besides, the model fit may be tested using the procedure described by \citeA{dasylva_goussanou_jjsds_2024}, to account for the correlation of the $n_i$'s.
 When $S_A$ is a census, the methodology may also serve to estimate the false negatives generated by the blocking procedure \cite{dasylva_goussanou_2021}.
 However, it must be adapted when $S_A$ has some undercoverage.


\section{Methodology}

The proposed methodology is based on two extensions of the model described in the previous section, which is subsequently called the {\it univariate neighbor model} or simply neighbor model.
 The first extension accounts for the undercoverage of $S_A$ and only  changes the interpretation of some model parameters.
 It is used to estimate the coverage when linking with a sufficiently high recall.
 The second extension is more substantial, as it replaces $n_i$ by a vector of such variables; each representing the number of links for a distinct linkage rule.
 The resulting model is called the {\it multivariate neighbor model}, which is used to estimate the coverage by specifying the interactions in the matched pairs, while allowing arbitrary interactions in the unmatched pairs.
 The following paragraphs discuss the linkage strategy before delving into the details of the various extensions and how they are used to estimate the coverage.


\subsection{Linking the sources}

In their solutions, \citeA{ding_fienberg_1994}, \citeA{diconsiglio_tuoto_2015} and \citeA{dewolf_et_al_2019} constrain each record to have at most one link.
 Yet, this constraint greatly complicates the modeling of the linkage errors as mentioned before.
 Here, it is instead proposed to link the records without this constraint,  with a rule such that the decision to link two records involves no other record.
 Thus, a record may have zero, one or many links.
 Such a linkage rule may be implemented with the deterministic, hierarchical or probabilistic methods of record linkage.


\subsection{Extending the univariate neighbor model}

When $S_A$ is not a census, the connection between $n_i$ and the errors is according to Table~\ref{table: connection between n_i and the linkage errors for a sample}, which differs from Table~\ref{table: connection between n_i and the linkage errors for a census} when $n_i=0$ and $n_i=|S_A|$.
 When $n_i=0$, there is no certainty about the occurrence of a false negative because it is not known if the corresponding unit is in $S_A$, unlike what happens in Table~\ref{table: connection between n_i and the linkage errors for a census}.
 When $n_i=|S_A|$, the number of false positives is not known with certainty for the same reason.
%
%
%
\begin{table}[htbp]
\centering
\caption{Connection between $n_i$ and errors when $S_A$ is not a census. \label{table: connection between n_i and the linkage errors for a sample}}
\vspace{6pt}
{\small
\begin{tabular}{ccc}\hline
$n_i$			  & False negatives & False positives \\ \hline
0			  & 0 or 1	    & 0 \\
$1 \leq n_i \leq |S_A|-1$  & 0 or 1 & $n_i-1$ or $n_i$ \\
$n_i=|S_A|$  & 0	& $|S_A|-1$ or $|S_A|$ \\ \hline
\end{tabular}}
\end{table}
To account for the undercoverage of $S_A$, redefine ${\cal B}_N(v)$ as a subset of ${\cal V}_N$ such that
%
%
\begin{equation}
 \label{eq: B_N(v)}
 {\cal B}_N(v) \supset
 \left \{ v' \in {\cal V}_N \ s.t. \
 E \left [ L_{ij} \left |
 (i,j) \in S_B \times S_A, \left ( V_i,V_j'\right )=(v,v') \right . \right ]>0
 \right \}
\end{equation}
 Also redefine $p_N(.)$, $\lambda_N(.)$ and $\lambda_N^{(0)}(.)$ as
%
%
\begin{eqnarray}
 \label{eq: p_N(v)}
 p_N(v) &=&
 E \left [ I \left ( i \in S_A \right ) L_{ii} \left |
 i \in S_B, V_i=v \right . \right ],\\
%
%
 \label{eq: lambda_N(v)}
 \lambda_N(v) &=&
 E \left [ I \left (j \in S_A \right ) L_{ij} \left |
 i \in S_B, V_i=v \right . \right ], \ j \neq i,\\
%
%
 \label{eq: lambda_N^{(0)}(v)}
 \lambda_N^{(0)}(v) &=&
 P \left ( j \in S_A,
 V_j' \in {\cal B}_N \left ( V_i \right ) \left |
 i \in S_B, V_i=v \right . \right ),
\end{eqnarray}
 where $p_N(v)$ is the joint probability of including $i$ in $S_A$ and having a true positive, $\lambda_N(v)$ is still the false positive probability and $\lambda_N^{(0)}(v)$ is the probability of having an unmatched record in the neighborhood.
 With these updated definitions, it is easily shown that Eq.~\ref{eq: limiting mixture} applies, when $N \rightarrow \infty$ under the regularity conditions given by Eqs.~\ref{eq: census regularity condition 1}-\ref{eq: census regularity condition 2} and $F$ is discrete with $G$ atoms.
 Indeed, the proof for the census case still applies with $n_i$ based on Eq.\ref{eq.: general n_i}. See \citeA[Lemma 1]{dasylva_goussanou_jjsds_2022} for the details.
 The parameters $\alpha_g$ and $\lambda_g$ have the same interpretation, but $p_g$ now corresponds to the product of the inclusion probability $P(i \in S_A)$ by the probability of a true positive for a record in class $g$.
 As before, let $\overline{p}=\sum_{g=1}^G \alpha_g p_g$ and $\overline{\lambda}=\sum_{g=1}^G \alpha_g \lambda_g$, where $\overline{p}$ is the expected number of true positives per record, which is also equal to $E \left [ \left . I(i \in S_A) L_{ii} \right | i \in S_B\right ]$, and $\overline{\lambda}$ is the expected number of false positives per record.
 Thus $\overline{p}$ is a useful lower-bound on $P(i \in S_A)$.
 The model parameters may be estimated by maximizing the composite likelihood of the $n_i$'s when $G$ is given, and by selecting this latter parameter through the minimization of Akaike's information criterion as in the census case \cite{dasylva_goussanou_jjsds_2022}.
 Let $\widehat{\overline{p}}$ and $\widehat{\overline{\lambda}}$ denote the resulting maximum likelihood estimators.
 In Appendix~\ref{appendix: undercoverage}, it is shown that the recall and precision (two finite population parameters) converge in probability to $P(i \in S_A)^{-1} \overline{p}=E \left [ \left . L_{ii} \right | i \in S_A \cap S_B\right ]$ and $\overline{p}/(\overline{p}+\overline{\lambda})$, respectively, under the regularity conditions.
 Under the same conditions, it is also shown that $\widehat{\overline{p}}$ and $\widehat{\overline{\lambda}}$ are consistent estimators of $\overline{p}$ and $\overline{\lambda}$, so that $P(i \in S_A)^{-1} \widehat{\overline{p}}$ and $\widehat{\overline{p}}/(\widehat{\overline{p}}+\widehat{\overline{\lambda}})$ are consistent estimators of the recall and precision, respectively.


\subsection{Estimating the coverage with a high recall}

From the above discussion, it follows that a consistent estimator of the coverage $P(i \in S_A)$ may be obtained as
 \[ \left ( \frac{TP}{TP+FN}\right )^{-1} \widehat{\overline{p}}, \]
 if the recall (i.e., $TP/(TP+FN)$) is known.
 In particular, $\widehat{\overline{p}}$ is consistent, if the recall is known to be perfect, i.e., $TP/(TP+FN)=1.0$, which is equivalent to having no false negatives.
 However, it must be noted that the proposed model is of interest only where the linkage is not perfect, i.e., if the precision is smaller than 1.0.
 Otherwise, the standard capture-recapture estimator would apply, including in the ideal situation where the linkage key is an error-free unique identifier, i.e., a perfect linkage key.
 Besides, the neighbor model is not advised with such a key, because some of the model assumptions may not hold.

To use the above approach in practice, one would want to design the linkage rule such that it generates very few false negatives if any, and ideally with a false negative rate smaller than $\min \left ( P(i \in S_A), 1-P(i \in S_A) \right )$ by an order of magnitude.
 This may be inspired from blocking procedures, which are used in probabilistic linkages to select a small subset of the Cartesian product with the majority of the matched pairs.
 \citeA{christen_2012} provides a good review of these procedures, which are indispensable when the sources are large.
 However, achieving a sufficiently high recall may come at the expense of tolerating a very low precision, which can prevent the estimation of the coverage with the required accuracy.
 In such cases, it is proposed to estimate the coverage by specifying the interactions in the matched pairs.
 However, this requires a multivariate extension of the neighbor model.


\subsection{Multivariate neighbor model}

The multivariate extension concerns finite collections of simple linkage rules that are also mutually exclusive, i.e., for each rule, the decision to link two records involves no other records, and each pair is linked by at most one rule.
 The need for this extension is best explained with an example.
 For simplicity, suppose that $S_A$ is known to be a complete census and that the two sources are to be linked with the given name, last name and birth date.
 To do so, seven rules are to be evaluated, which are shown in Table~\ref{table: seven exclusive rules example}, where $\gamma$ lives in the finite set $\Gamma = \{0,1\}^3 \setminus \{(0,0,0)\}$, and
 the components of $\gamma$ indicate whether there is an exact agreement on the last name, given name and birth date, in this order.
 For rule $\gamma$, denote by $n_i^{(\gamma)}$ the corresponding number of links for the sample record $i$, e.g., $n_i^{(0,0,1)}$ is the number of links when linking based on having the same names but a different birth date.
%
%
\begin{table}[htbp]
\begin{center}
\caption{Mutually exclusive rules based on the given name, last name and birth date.\label{table: seven exclusive rules example}}\vspace{6pt}
%
{\small
\begin{tabular}{cccc}\hline
 Rule index $\gamma=(\gamma_1,\gamma_2,\gamma_3)$ & Same last name & Same given name & Same birth date \\ \hline
 $(0,0,1)$ & \xmark & \xmark & \cmark \\
 $(0,1,0)$ & \xmark & \cmark & \xmark \\
 $(0,1,1)$ & \xmark & \cmark & \cmark \\
 $(1,0,0)$ & \cmark & \xmark & \xmark \\
 $(1,0,1)$ & \cmark & \xmark & \cmark \\
 $(1,1,0)$ & \cmark & \cmark & \xmark \\
 $(1,1,1)$ & \cmark & \cmark & \cmark \\ \hline
\end{tabular}}
\end{center}
\end{table}
 A simple way to evaluate the different rules is to fit a model of the form
%
%
\begin{equation}
 \label{eq: n_i_gamma distribution}
 \left . n_i^{(\gamma)} \right | \{i \in S_B\} \sim
 \sum_{g=1}^{G^{(\gamma)}} \alpha_g^{(\gamma)}
 Bernoulli \left ( p_g^{(\gamma)} \right ) \ast
 Poisson \left ( \lambda_g^{(\gamma)} \right ),
\end{equation}
 separately for each $\gamma$, where the expected numbers of true positives and false positives per record are $ \overline{p}^{(\gamma)} = \sum_{g=1}^{G^{(\gamma)}} \alpha_g^{(\gamma)} p_g^{(\gamma)}$ and $\overline{\lambda}^{(\gamma)} = \sum_{g=1}^{G^{(\gamma)}} \alpha_g^{(\gamma)} \lambda_g^{(\gamma)}$, respectively.
 Note that we necessarily have the constraint $\sum_{\gamma \in \Gamma} \overline{p}^{(\gamma)} \leq 1$, because the rules are mutually exclusive.
 However, the resulting estimators of the recall and precision may be inefficient because important information is ignored, such as the constraint $\sum_{\gamma \in \Gamma} \overline{p}^{(\gamma)} \leq 1$, or the correlation among the numbers of links from the different rules for the same sample record, which is not exploited either.
 Also, when choosing the number of classes $G^{(\gamma)}$ according to Akaike's information criterion, the result $\widehat{G}^{(\gamma)}$ may vary across the different rules, which is counterintuitive.
 Besides, even in the best case where $\widehat{G}^{(\gamma)}$ is the same for all the rules, the classes may correspond to different  latent partitions of the sample records across the different rules, which is also counterintuitive and undesirable.
 Furthermore, The above limitations apply in the more general situation where $S_A$ is not a census and its coverage is unknown.

The solution is to model the joint distribution of the vector of counts $\left [ n_i^{(\gamma)}\right ]_{\gamma \in \Gamma}$ with a multivariate extension of the neighbor model, as follows, with further details in Appendix~\ref{appendix: multivariate}.
 To describe this extension, it is convenient to define the following multivariate distributions.
 The first distribution is the joint distribution of mutually independent variables that are indexed over a finite set $\Gamma$, where variable $\gamma$ follows the $Poisson \left ( \lambda^{(\gamma)} \right )$ distribution.
 Thus, the joint distribution is simply the product distribution.
 For notational convenience, we define $\bm{\lambda} = \left [ \lambda^{(\gamma)}\right ]_{\gamma \in \Gamma}$ and denote this distribution by $PPoisson \left ( \bm{\lambda} \right )$, where the first ``P'' stands for product.
 The second distribution corresponds to the joint distribution of the cell counts in a multinomial experiment with $n$ trials, where the last cell is excluded, the other cells are indexed over $\Gamma$, and the probability of observing cell $\gamma$ is denoted by $p^{(\gamma)}$, such that $\sum_{\gamma \in \Gamma} p^{(\gamma)} \leq 1$.
 In this case, we define $\bm{p} = \left [ p^{(\gamma)}\right ]_{\gamma \in \Gamma}$ and denote the joint distribution by $IMultinomial \left (n, \bm{p} \right )$, where the ``I'' stands for incomplete.
 In general, the multivariate extension may be considered for modeling the joint distribution of the numbers of links, which result from the application of mutually exclusive simple linkage rules that are indexed over some finite set $\Gamma$, where $n_i^{(\gamma)}$ denotes the number of links from rule $\gamma$ for the sample record $i$ and $\bm{n}_i = \left [ n_i^{(\gamma)}\right ]_{\gamma \in \Gamma}$.
 The multivariate model is a finite mixture of $|\Gamma|$-variate discrete distributions, where each component is the convolution of an incomplete multinomial distribution with a product of independent Poisson distributions, i.e.,
%
%
\begin{equation}
\label{eq: multivariate neighbor}
 \left . \bm{n}_i \right | \{i \in S_B\} \sim \sum_{g=1}^G
 \alpha_g IMultinomial(1,\bm{p}_g) \ast
 PPoisson \left ( \bm{\lambda}_g \right ),
\end{equation}
 where $G$ is the number of record classes, $\alpha_g$ is the probability that a sample record is from class $g$, and $\bm{p}_g=\left [ p_g^{(\gamma)} \right ]_{\gamma \in \Gamma}$ and
 $\bm{\lambda}_g= \left [ \lambda_g^{(\gamma)} \right ]_{\gamma \in \Gamma}$ are the vectors of expected numbers of true positives and false positives for a record in the class.
 Furthermore, $p_g^{(\gamma)}$ is the expected number of true positives and $\lambda_g^{(\gamma)}$ is the expected number of false positives, under rule $\gamma$.
 Then, given $G$, the model is parametrized by $\left [ \left ( \alpha_g, \bm{p}_g, \bm{\lambda}_g \right ) \right ]_{1 \leq g \leq G}$.
 When the records are homogeneous,
%
%
\begin{equation}
 \left . \bm{n}_i \right | \{i \in S_B\} \sim
 IMultinomial(1,\bm{p}) \ast
 PPoisson \left ( \bm{\lambda} \right ),
\end{equation}
 where $\bm{p} = \left [ p^{(\gamma)} \right ]_{\gamma \in \Gamma}$ and $\bm{\lambda} = \left [ \lambda^{(\gamma)}\right ]_{\gamma \in \Gamma}$.
 Furthermore, if $\min_{\gamma \in \Gamma} \lambda^{(\gamma)} >0$ and $\bm{t} = \left [ t^{(\gamma)} \right ]_{\gamma \in \Gamma}$, we have
%
%
\begin{eqnarray}
 P \left ( \left . \bm{n}_i = \bm{t} \right | i \in S_B \right )
 &=&
 I \left ( |\bm{t}|=0 \right )
 \left ( 1- |\bm{p}|\right ) e^{-|\bm{\lambda}|} + \nonumber \\
 & &
 I \left ( |\bm{t}|>1 \right )
 \left ( \left ( 1- |\bm{p}| \right )
 \prod_{\gamma \in \Gamma}
 \frac{e^{-\lambda^{(\gamma)}}
 \left ( \lambda^{(\gamma)}\right )^{t^{(\gamma)}}
 }{t^{(\gamma)}!} \right . + \nonumber \\
 & &
 \left . 
 \sum_{\gamma \in \Gamma: \ t^{(\gamma)}>0} p^{(\gamma)}
 \frac{e^{-\lambda^{(\gamma)}}
 \left ( \lambda^{(\gamma)}\right )^{t^{(\gamma)}-1}
 }{\left ( t^{(\gamma)} -1\right )!}
 \prod_{\gamma' \in \Gamma \setminus \{\gamma \}}
 \frac{e^{-\lambda^{(\gamma')}}
 \left ( \lambda^{(\gamma')}\right )^{t^{(\gamma')}}
 }{t^{(\gamma')}!} \right ).
\end{eqnarray}
 Like before, the model is motivated by the convergence in distribution of the vector of counts $\bm{n}_i=\left [ n_i^{(\gamma)} \right ]_{\gamma \in \Gamma}$, when $N \rightarrow \infty$, as stated by Lemma~\ref{lemma: multivariate convergence in distribution} in Appendix~\ref{appendix: multivariate}.
 From the multivariate mixture, it follows that the marginal distribution of $n_i^{(\gamma)}$ is still given by Eq.~\ref{eq: n_i_gamma distribution}, except that $G^{(\gamma)}$ and $\alpha_g^{(\gamma)}$ are the same for all the rules, as desired.
 Also, the record classes correspond across all the rules now.
 A restricted model is obtained when $\bm{p}_g = \varrho \left ( \bm{\beta}_g \right )$ for each class, where $\varrho(.)$ is a known injective function and $\bm{\beta}_g$ is a vector of regression coefficients of dimension smaller than $|\Gamma|$.
 This means that the $|\Gamma|$ true positive probabilities for the different rules are not free but bound by fewer regression coefficients, for each record class.
 Such a restriction is useful when exploiting the information about the variables'interactions in the matched pairs, through a log-linear specification.
%
%
 Then, a multivariate mixture with $G$ components is parametrized by $\left [ \left ( \alpha_g, \bm{\beta}_g, \bm{\lambda}_g\right ) \right ]_{1 \leq g \leq G}$.
 According to Theorem~\ref{theorem: multivariate consistent estimation} in Appendix~\ref{appendix: multivariate}, the model parameters (for each proposed parameterization) is estimated consistently by maximizing the composite likelihood of the observed vectors of counts $\left [ \bm{n}_i\right ]_{i \in S_B}$, when $G$ is known or unknown.
 In the latter case, $G$ may be selected according to the minimum Akaike's information criterion as before.


\subsection{Estimating the coverage through the correlation structure} \label{subsection: exploiting the correlation structure}

When the records are linked with a perfect recall, the coverage may be estimated with the univariate mixture model as discussed before.
 Otherwise, the coverage may be estimated with the multivariate neighbor model, where the true positive probabilities are constrained according to the correlation structure of the linkage variables through a log-linear specification.
 In detail, the proposed multivariate model is based on a collection of simple linkage rules (i.e., each rule is such that the decision to link two records involves no other record), which are themselves based on a first set of simple linkage rules that are divided into $K$ groups.
 In group $k$, there are $H_k$ mutually exclusive rules (i.e., a pair is linked by at most one rule from the group) and $L_{ij}^{(k,h)}$ indicates whether the pair $(i,j)$ is linked by rule $h$.
 For example, each group may be based on a single variable, and the rules may correspond to different levels of agreement on this variable.
 For example, for the last name, these levels may comprise exact agreement, typo agreement (excluding exact agreement) and SOUNDEX agreement (excluding exact and typo agreements).
 However, a group may also involve many variables.
 A second set of rules is obtained by combining the rules from the first set as follows.
 Let the index set be $\Gamma =
  \{0,\ldots,H_1\}\times \ldots \times \{0,\ldots,H_K\}-\bm{0}_K$, and for $\gamma = (\gamma_1,\ldots,\gamma_K) \in \Gamma$, let $L_{ij}^{(\gamma)}$ denote the indicator that rule $\gamma$ links the pair $(i,j)$, where $L_{ij}^{(\gamma)}=1$ only if $L_{ij}^{(k,\gamma_k)} = 1$ for each $k$ such that $\gamma_k \geq 1$, and $\sum_{h=1}^{H_k} L_{ij}^{(k,h)}=0$ for each $k$ such that $\gamma_k=0$.
 The proposed model is the special case of the multivariate neighbor model (Eq.~\ref{eq: multivariate neighbor}), where $p_g^{(\gamma)}$ has the following form, for a vector of covariates $\bm{z}^{(\gamma)}$ and regression coefficients $\bm{u}_g$.
%
%
\begin{equation}
 \label{eq: restricted multivariate neighbor p_g}
 p_g^{(\gamma)} =
 \frac{P(i \in S_A)
 \exp \left ( {\bm{z}^{(\gamma)}}^{\top} \bm{u}_g
 \right )}{1+
 \sum_{\gamma' \in \Gamma}
 \exp \left ( {\bm{z}^{(\gamma')}}^{\top} \bm{u}_g
 \right )}.
\end{equation}
 For a given number of classes $G$, the model parameters include  $\left [ \left ( \alpha_g, \bm{u}_g, \bm{\lambda_g} \right ) \right ]_{1 \leq g \leq G}$ and $P(i \in S_A)$.
 The specific form of $\bm{z}^{(\gamma)}$ and $\bm{u}_g$ depends on the model.
 This is illustrated in the following example, where the model includes all the main terms and no interactions.
 This is also saying that the components of $\gamma$ are independent in the matched pairs.
 In this case, the coefficient corresponding to the event $\{ \gamma_k = l_k \}$ is denoted by $u_{g,k \left ( l_k \right )}$.
 By the dummy coding convention, the coefficient is set to zero if $l_k=0$.
 The covariate corresponding to this coefficient is the indicator $I(\gamma_k=l_k)$ so that
%
%
\begin{eqnarray}
 \label{eq: multivariate neighbor no interaction z}
 \bm{z}^{(\gamma)} &=&
 \left [ I \left ( \gamma_1=1 \right ) \ldots
 I \left ( \gamma_1=H_1 \right ) \ldots
 I \left ( \gamma_K=1 \right ) \ldots
 I \left ( \gamma_K=H_K \right ) \right ]^{\top},\\
 \label{eq: multivariate neighbor no interaction u}
 \bm{u}_g &=&
 \left [ u_{g,1(1)} \ldots u_{g,1(H_1)} \ldots
         u_{g,K(1)} \ldots u_{g,K(H_K)} \right ]^{\top}.
\end{eqnarray}
 In the next example, the model includes all the main terms and second-order interactions, but no higher-order interactions.
 For $1 \leq k_1 < k_2 \leq K$, the coefficient of the interaction between the events $\{\gamma_{k_1} = l_{k_1}\}$ and $\{\gamma_{k_2} = l_{k_2}\}$ is denoted by $u_{g,k_1 k_2 \left ( l_{k_1}l_{k_2} \right )}$.
 By the same coding convention, the coefficient is set to zero if $l_{k_1}=0$ or $l_{k_2}=0$.
 The covariate associated with the coefficient is the indicator $I \left ( \left ( \gamma_{k_1},\gamma_{k_2}\right )=\left ( l_{k_1},l_{k_2}\right ) \right )$.
 In this case, $\bm{z}^{(\gamma)}$ and $\bm{u}_g$ have more complex expressions.
 In order to write them in a manner that is tidy, define
%
%
\begin{equation*}
 \bm{z}^{(\gamma)}_{1k} =
 \left [ I \left ( \gamma_k=1 \right ) \ldots
 I \left ( \gamma_k=H_k \right ) \right ]^{\top}.
\end{equation*}
 Also, denote the right-hand side of Eq.~\ref{eq: multivariate neighbor no interaction z}
 by $\bm{z}^{(\gamma)}_1$, the right-hand side of
 Eq.~\ref{eq: multivariate neighbor no interaction u} by $\bm{u}^{(\gamma)}_{g1}$, and, for $k=1,\ldots,K-1$, further define
%
%
\begin{eqnarray*}
 \bm{z}^{(\gamma)}_{2k} &=&
 \left ( \left [ {{} \bm{z}_{1(k+1)}^{(\gamma)}}^{\top} \ldots {{} \bm{z}_{1K}^{(\gamma)}}^{\top} \right ] \otimes
 {{} \bm{z}_{1k}^{(\gamma)}}^{\top} \right )^{\top}, \\
 \bm{u}_{g2k} &=&
 \Big [ \overbrace{u_{g,k(k+1)(11)} \ldots u_{g,k (k+1) (H_k 1)} }^{
 \scriptsize
 \begin{array}{l}
 interaction \ terms \ between\\
 level \ 1 \ of \ \gamma_{k+1} \ and \\
 all \ levels \ of \ \gamma_k
 \end{array}} \ldots
 \overbrace{u_{g,k(k+1)(1H_{k+1})} \ldots
 u_{g,k (k+1) (H_k H_{k+1})}}^{
 \scriptsize
 \begin{array}{l}
 interaction \ terms \ between\\
 level \ H_{k+1} \ of \ \gamma_{k+1} \ and \\
 all \ levels \ of \ \gamma_k
 \end{array}}
 \ldots \nonumber \\
 & &
 \overbrace{u_{g,k K (11)} \ldots u_{g,k K (H_k 1)}}^{
 \scriptsize
 \begin{array}{l}
 interaction \ terms \ between\\
 level \ 1 \ of \ \gamma_K \ and \\
 all \ levels \ of \ \gamma_k
 \end{array}} \ldots
 \overbrace{u_{g,k K (1H_K)} \ldots u_{g,k K (H_k H_K)}}^{
 \scriptsize
 \begin{array}{l}
 interaction \ terms \ between\\
 level \ H_K \ of \ \gamma_K \ and \\
 all \ levels \ of \ \gamma_k
 \end{array}}
 \Big ]^{\top},
\end{eqnarray*}
 where $\otimes$ is the Kronecker product and $\bm{u}_{g2k}$ are the interaction terms between $\gamma_k$ and $\gamma_{k+1},\ldots,\gamma_K$.
 Then, we have
%
%
\begin{eqnarray*}
 \bm{z}^{(\gamma)} &=&
 \left [ {\bm{z}^{(\gamma)}_1}^{\top} \ 
         {\bm{z}^{(\gamma)}_{21}}^{\top} \ldots
         {\bm{z}^{(\gamma)}_{2(K-1)}}^{\top} \right ]^{\top},\\
 \bm{u}_g &=&
 \left [ \bm{u}_{g1}^{\top} \ 
         \bm{u}_{g21}^{\top} \ldots
         \bm{u}_{g2(K-1)}^{\top} \right ]^{\top}.
\end{eqnarray*}
In general, for $t\leq K$, the coefficient of the interaction between the events $\{\gamma_{k_1} = l_{k_1}\}$,...,$\{\gamma_{k_t} = l_{k_t}\}$ is denoted by $u_{g,k_1\ldots k_t \left ( l_{k_1} \ldots l_{k_t} \right )}$.
 As before, the coefficient is set to zero by convention if $\min \left ( l_1,\ldots,l_t\right )=0$.
 The covariate associated with the coefficient is the indicator $I \left ( \left ( \gamma_{k_1},\ldots,\gamma_{k_t}\right )=\left ( l_{k_1},\ldots,l_{k_t}\right ) \right )$.
 When the model includes all the main terms and interactions of order $d$ or smaller, we have
%
%
\begin{eqnarray*}
 {\bm{z}^{(\gamma)}}^{\top} \bm{u}_g &=&
 \sum_{t=1}^d \sum_{1 \leq k_1<\ldots<k_t \leq K}
 \sum_{l_{k_1}=1}^{H_{k_1}} \ldots
 \sum_{l_{k_t}=1}^{H_{k_t}} \\
 & &
 I \left ( \left ( \gamma_{k_1},\ldots ,\gamma_{k_t} \right ) =
 \left ( l_{k_1},\ldots l_{k_t}, \right ) \right ) 
 u_{g,k_1 \ldots k_t \left ( l_{k_1}\ldots l_{k_t} \right )}.
\end{eqnarray*}

According to Theorem~\ref{theorem: multivariate consistent estimation} in Appendix~\ref{appendix: multivariate}, this implies that the parameters of the limiting mixture (including the coverage $P(i \in S_A)$) can be estimated consistently by maximizing the likelihood of the $\bm{n}_i$'s, under the stated conditions.
 For example, this methodology may be of interest in the following simple setup, where the linkage is based on exact comparisons of the last name (first group), given name (second group) and birth date (third group), with $K=3$, $H_1=H_2=H_3=1$, $\Gamma = \{0,1\}^3 \setminus \{(0,0,0)\}$ and $|\Gamma|=7$.
 The coverage may be estimated by maximizing the likelihood of the $\bm{n}_i$'s, if the different agreements have no interactions of the third order in matched pairs for each possible value of a record in $S_B$, i.e., all main terms and second order interactions may be included (this adds up to six parameters besides the coverage $P(i \in S_A)$).
 In particular, this is true if the agreements are independent in matched pairs.
 In this case, the solution is related to that described by \citeA{racinskij_et_al_2019}.
 It is also related to the solution described by \citeA{brown_et_al_2020} except that it doest not resort to clerical reviews.
 Beyond this special case, the true positives distribution is associated with seven unknown parameters ($P(i \in S_A)$ and the six log-linear parameters) and seven equations (one for $p^{(\gamma)}$ for each $\gamma$), for each mixture component.

 A few remarks are in order.
 The first remark is that the proposed model implicitly accounts for all the interactions among the linkage variables in the unmatched pairs, while accounting for all the interactions of order $K-1$ or smaller in the matched pairs, within each record class.
 Thus, it offers a far greater modeling flexibility than classical log-linear mixtures, while retaining the identification property (see Lemmas~\ref{lemma: identification multivariate} and \ref{lemma: injective log-linear}) and the ability to consistently estimate the related parameters (see Theorem~\ref{theorem: multivariate consistent estimation}).
 This is best seen in the simpler case where the true positives distribution is homogeneous across the records and $H_1=\ldots=H_K=1$, i.e., $K$ dichotomous comparisons.
 In this case, it is clear that one cannot use a two-component $K$-variate log-linear mixture to model the record pairs, while accounting for all the interactions of order $K-1$ or smaller in the matched pairs.
 Indeed, this entails having at least $2^K$ free parameters, including $2^K-2$ parameters for the matched pairs, one parameter for the mixing proportion and at least one parameter for the unmatched pairs.
 However, there are only $2^K$ observable patterns and thus only $2^K-1$ equations to determine the parameters.
 The same problem occurs when $K=2$, even when there are no interactions in the distribution of the matched pairs.
 The second remark is that the added modeling flexibility greatly facilitates the design of linkage rules, which meet the conditions for the consistent estimation of the coverage (see Theorem~\ref{theorem: multivariate consistent estimation} and Lemma~\ref{lemma: injective log-linear}).


\subsection{Heterogeneous capture and incomplete records}

The methodology may be applied when the capture probability varies across post-strata according to covariates that are recorded without errors in each sample, so long as the stated assumptions (see Sections~\ref{subsection: finite population and sources}, \ref{subsection: model for heterogeneous records}
 and the appendix) hold within each post-stratum.
 In this case, $S_A$ and $S_B$ correspond to the subsets of records from a post-stratum, where the capture probability may be estimated using one of the neighbor models.
 Of course, the construction of the post-strata is an important practical question, which is deferred to future work.

Another practical concern is the occurrence of incomplete records in either sample.
 To discuss the issue without burdening the notation, let $S_A$ and $S_B$ now denote the two samples within a post-stratum, let $S_A'$ and $S_B'$ denote the corresponding subsamples of complete records, and let $P(i \in S_A')$ denote the coverage of $S_A'$ within the post-stratum, where $i$ is a unit located therein.
 \citeA[chap. 1.2]{little_rubin_1987} have described different strategies for conducting a statistical analysis in the presence of incomplete records.
 A first option is to use only the complete records, with or without weighting them to account for the incomplete ones.
 Two other options include imputing the missing values and the model-based approach, which consists in maximizing the likelihood of the incomplete data.
 Here, the first option may be considered without reweighting the complete records, when the stated assumptions apply within each post-stratum, including the fact that the inclusion in $S_A'$ and that in $S_B'$ are independent and the inclusion probability in $S_A'$ is uniform.
 The main idea is to treat the missingness as a second stage of selection and estimate the coverage in two steps as follows.
 First, apply one of the two proposed methodologies within the post-stratum to obtain an estimate $\widehat{P}(i \in S_A')$ of the coverage of $S_A'$.
 Second, estimate the coverage of $S_A$ (within the post-stratum) by $|S_A|/(|S_A'|/\widehat{P}(i \in S_A'))$.


\section{Simulations}

The proposed methodology is evaluated with Monte Carlo simulations comprising 100 repetitions.
 In a repetition, a finite population is generated with 100,000 individuals, where each individual is assigned a surname and birth date.
 From this population two Bernoulli samples are drawn, where the surname and birth date are possibly recorded with typos.
 Then, the samples are linked and the coverage is estimated with the proposed models and according to \citeA{ding_fienberg_1994}, \citeA{diconsiglio_tuoto_2015} and \citeA{racinskij_et_al_2019}, for comparison.
 Different scenarios are considered with different linkage rules, including some where the conditional independence assumption applies and the recall is perfect. The following paragraphs provide more details.


\subsection{Finite population and data sources}

For the surname and birth date, the frequencies are based on crossing the surname and age distributions from the 2010 US census of population \cite{uscb_2010_2019_pop,uscb_2010_censuS_surnames}.
 For the surname, the relative frequency is computed after excluding the observation ``all other surnames''.
 For simplicity, the month is uniformly drawn from $\{1,\ldots,12\}$ and the day is independently and uniformly drawn from $\{1,\ldots,30\}$.
 Consequently, the surname and date components are mutually independent in the population.
 Two complete registers are created, where the variables are perturbed in the second register.
This perturbation is described in terms of exact agreement on the surname, birth day, or birth month, and a baseline criterion, which is defined as having the same surname SOUNDEX and birth year, as well as an absolute difference that is smaller than 2 for both the day and the month.
 To be more specific, let $\gamma_1$, $\gamma_2$ and $\gamma_3$ denote the indicator variables, which correspond to the satisfaction of the baseline criterion in addition to having the same surname, birth day or birth month, as shown in Table~\ref{table: simulation indicator variables}.
 For example, when $\gamma_1 = 1$, the baseline criterion is satisfied and the surname is identical.
 When $\gamma_1 = 0$, the baseline criterion is not met or it is met and the surname is different.
 In each case, the birth day and birth month may be identical or different.
%
%
\begin{table}[htbp]
\centering
\caption{Indicators of the perturbations in a record. \label{table: simulation indicator variables}}
\vspace{6pt}
{\small
\begin{tabular}{ccccc}\hline
Indicator	&	Baseline criterion	&	Same surname	&	Same birth day	&	Same birth month	\\ \hline
$\gamma_1$	&	\cmark	&	\cmark	&	?	&	?	\\
$\gamma_2$	&	\cmark	&	?	&	\cmark	&	?	\\
$\gamma_3$	&	\cmark	&	?	&	?	&	\cmark	\\ \hline
\end{tabular}}
\end{table}
 For a given individual, the related records in the first and second registers are hereafter called first and second records, respectively.
 The second record is obtained by first drawing $\gamma=(\gamma_1,\gamma_2,\gamma_3)$, and then choosing the record value according to the value of the first record and $\gamma$.
 For example, when $\gamma=(0,1,1)$, the second record is such that it has the same birth date as the first record but a different surname with the same SOUNDEX code.
 Using the dummy coding convention, we can write the distribution of $\gamma$ in log-linear form as
%
%
\begin{equation}
 \label{eq: simulations gamma distribution}
 P(\gamma) =
 \exp \left ( u + \sum_{k=1}^3 \gamma_k u_{k(1)}+
 \sum_{1 \leq k_1 < k_2 \leq 3}
 \gamma_{k_1} \gamma_{k_2} u_{k_1k_2(11)} +
 \gamma_1 \gamma_2 \gamma_3 u_{123(111)} \right ),
\end{equation}
 where the intercept $u$ is a function of the main terms and the interaction terms because $\sum_{\gamma \in \{0,1\}^3} P(\gamma) = 1$.
 For simplicity, we choose the parameters such that $u_{1(1)}=u_{2(1)}=u_{3(1)}$ and $u_{12(11)}=u_{13(11)}=u_{23(11)}$.
 When $\gamma_1=0$, the surname in the second record is drawn from the other census surnames with the same SOUNDEX code, according to their frequencies.
 When $\gamma_2=0$, the day in the second record is obtained by randomly increasing or decreasing the day of the first record by 1, with probability $1/2$ for each alternative, except when the day is $1$ or $30$ in the first record, in which case the day is increased by $1$ or decreased by $1$, respectively.
 Likewise, when $\gamma_3=0$, the month in the second record is obtained by randomly increasing or decreasing the month of the first record by 1, with probability $1/2$ for each alternative, except when the month is $1$ or $12$ in the first record, in which case the month is increased by $1$ or decreased by $1$, respectively.
 From each register, an independent Bernoulli sample is drawn with an inclusion probability of 0.9, which is the actual coverage.


\subsection{Linkage}

Two linkage rules are considered, where the first rule links the pairs that meet the baseline criterion, or the subset of these pairs where there is at least one exact agreement on the surname, day of birth or month of birth, depending on the scenario.
 The resulting links are used to estimate the coverage with the univariate and multivariate neighbor models, and a classical log-linear mixture model that incorporates the conditional independence assumption as described by \citeA{racinskij_et_al_2019}.
 For a given pair, the vector of outcomes is based on the indicators of exact agreement for the surname, day of birth and month of birth, e.g., $(1,1,1)$ for a perfect agreement on the surname and the two date components.
 In order to estimate the coverage with the methodologies proposed by  \citeA{ding_fienberg_1994}, and \citeA{diconsiglio_tuoto_2015}, a second linkage rule is required, with at most one link per record as well as clerical estimates of the resulting linkage accuracy.
 This rule is derived from the first one by deleting a link, if at least one involved record has many links.
 The clerical estimates of the linkage accuracy are based on drawing a simple random sample of 1,000 record pairs, which satisfy the baseline criterion, and using the truth deck.
 Note this procedure ignores the false negatives generated by the baseline criterion (akin to a blocking criterion), where they exist.


\subsection{Scenarios}

Five scenarios are considered.
 In the first scenario, the conditional independence assumption is satisfied based on $u_{1(1)}=1$, $u_{12(11)}=0$ and $u_{123(111)}=0$, and the first linkage rule is based on the baseline criterion and it has a perfect recall.
 In the second scenario, there is a departure from conditional independence due to interactions of the second order, based on $u_{1(1)}=1$ and $u_{12(11)}=1$ and $u_{123(111)}=0$, but the first linkage rule is still based on the baseline criterion.
 In the third scenario, there is also a departure from conditional independence due to interactions of the second and third order, based on $u_{1(1)}=1$, $u_{12(11)}=1$ and $u_{123(111)}=1/4$, but there is no change to the first linkage rule.
 The fourth scenario is identical to the second scenario, except that 
 the first linkage rule now links the pairs that meet the baseline criterion and have at least one exact agreement on the surname, day of birth or month of birth.
 Finally, the fifth scenario is identical to the fourth scenario excepts that third order interaction is added based on $u_{123(111)}=1/4$.
 The characteristics of the different scenarios are summarized in Tables~\ref{table: simulation scenarios} and \ref{table: simulation error rates}.
 In the latter table, the figures are based on averages across the repetitions.
%
%
\begin{table}[htbp]
\centering
\caption{Simulation scenarios. \label{table: simulation scenarios}}
\vspace{12pt}
\resizebox{0.7\textwidth}{!}{
\begin{tabular}{ccccc}\hline
	&	\multicolumn{3}{c}{Log-linear parameters}					&		\\ \cline{2-4}
Scenario	&	$u_{k(1)}$	&	$u_{k_1 k_2 (11)}$	&	$u_{123(111)}$	&	Conditional independence	\\ \hline
1	&	$1$	&	$0$	&	$0$	&	\cmark	\\
2 \& 4	&	$1$	&	$1$	&	$0$	&	\xmark	\\
3 \& 5	&	$1$	&	$1$	&	$1/4$	&	\xmark	\\ \hline
\end{tabular}
}
\end{table}
%
%
\begin{table}[htbp]
\centering
\caption{Empirical averages of the rates of linkage error. \label{table: simulation error rates}}
\vspace{12pt}
\resizebox{0.9\textwidth}{!}{
\begin{tabular}{cccccc}\hline
Scenario	&	Linkage	&	Recall	&	Precision	&	False positive rate (FPR) $\times 10^{-9}$	&	Perfect recall	\\ \hline
\\[-6pt]											
1	&	1	&	1.000	&	0.952	&	498.92	&	\cmark	\\
	&	2	&	0.944	&	1.000	&	4.15	&	\xmark	\\
	&		&		&		&		&		\\
2 \& 3	&	1	&	1.000	&	0.953	&	495.03	&	\cmark	\\
	&	2	&	0.950	&	1.000	&	4.17	&	\xmark	\\
	&		&		&		&		&		\\
4 \& 5	&	1	&	0.996	&	0.964	&	371.84	&	\xmark	\\
	&	2	&	0.957	&	1.000	&	3.48	&	\xmark	\\ \hline
\end{tabular}}
\end{table}




\subsection{Estimators}

The neighbor models are applied under a homogeneous distribution of the true positives, to reflect the current setup and the situation in practice, where the heterogeneity of the false positives distribution is expected to be the dominant source of heterogeneity for the $n_i$ distribution \cite{dasylva_goussanou_jjsds_2022}.
 This means that the probability $p_g$ and the vector $\bm{p}_g = \left [ p_g^{(\gamma)}\right ]_{\gamma \in \Gamma}$ are the same for all the classes.
 It also means that the parameter $\bm{\beta}_g$ is the same across the classes, if $\bm{p}_g$ is of the form $\varrho \left ( \bm{\beta}_g \right )$ for some known function $\varrho(.)$.
 For convenience, let $p$, $\bm{p}$ and $\bm{\beta}$ denote the common values of $p_g$, $\bm{p}_g$ and $\bm{\beta}_g$, respectively.
 Also let
\begin{equation}
 \label{eq: postulated gamma distribution}
 r^{(\gamma)} =
 \exp \left ( u + \sum_{k=1}^3 \gamma_k u_{k(1)}+
 \sum_{1 \leq k_1 < k_2 \leq 3}
 \gamma_{k_1} \gamma_{k_2}
 u_{k_1k_2(11)}\right ), \ \gamma \in \Gamma = \{0,1\}^3,
\end{equation}
 where $\sum_{\gamma \in \{0,1\}^3} r^{(\gamma)}=1$, the intercept $u$ is a function of the other parameters, and the right-hand side only includes interactions of the second order unlike that of Eq.~\ref{eq: simulations gamma distribution}.
 Then
\begin{equation*}
 \bm{\beta} =
 \left (u_{1(1)},u_{2(1)},u_{3(1)},u_{12(11)},u_{13(11)},
 u_{23(11)} \right ),
\end{equation*}
 and the mapping $\varrho(.)$ is characterized by
\begin{equation*}
 p^{(\gamma)} =
 P(i \in S_A) r^{(\gamma)}, \
 \gamma \in \Gamma = \{0,1\}^3 \setminus \{(0,0,0)\}.
\end{equation*}
 The estimates are computed by maximizing the likelihood numerically in R, where the number of classes is chosen by minimizing Akaike's information criterion.
 For the univariate model, the estimates are based on capping the $n_i$'s by 10 (i.e., replacing $n_i$ by $min(10,n_i)$) and maximizing the likelihood of the resulting observations, as described by \citeA{dasylva_goussanou_jjsds_2022}.
 With the multivariate model, the coverage is estimated when only including the main terms, and when also including the second order interaction terms, while ignoring that the main terms are equal, and  that the second order interaction terms are equal.
 In general, the numerical maximization of the likelihood is more challenging than with the univariate model, and the resulting estimates become less accurate when the linkage precision decreases.
 Consequently, a good initialization procedure is needed, which is described in the Appendix~\ref{appendix: initialization}.
 For comparison, we also compute the estimators proposed by \citeA{ding_fienberg_1994}, \citeA{diconsiglio_tuoto_2015}, and \citeA{racinskij_et_al_2019}, as well as the naive capture-recapture estimator, which ignores the linkage errors.
 Note that this latter estimator is computed as the ratio of the number of links by the second linkage rule over $|S_B|$.


\subsection{Results}

The simulation results are shown in Table~\ref{table: simulation results}.
 In scenario 1, where the conditional independence assumption applies, the best performance is obtained with the estimators by \citeA{racinskij_et_al_2019} and the neighbor estimators, both in terms of the relative bias and the mean square error, with an advantage for the neighbor estimators when looking at this latter performance measure.
 Among the neighbor estimators, the univariate model offers the best  performance in terms of bias, variance and mean square error.
 Without surprise, the naive estimator has the worst performance, while the estimators by \citeA{ding_fienberg_1994}, and \citeA{diconsiglio_tuoto_2015} perform better but have a large variance, because they incorporate clerical estimates of the linkage accuracy.
 It is notable that the log-linear mixture proposed by \citeA{racinskij_et_al_2019} has a larger bias, variance and mean square error, than the estimators based on the multivariate neighbor models, with one small exception\footnote{In Table~\ref{table: simulation results}, the log-linear mixture has a relative bias slightly smaller that of the multivariate neighbor model with interactions.}.
 Indeed, all these estimators aim to estimate the coverage by leveraging the correlation structure of the linkage variables, which the log-linear mixture does fully by incorporating the independence of the linkage variables both in the matched pairs and in the unmatched pairs.
 However, the multivariate neighbor models only exploit the information about the correlation structure in the matched pairs without constraint on the unmatched ones.
 Yet, they yield estimators that are significantly more accurate than the log-linear mixture in terms of mean square error.
 This illustrates the important difference between classical log-linear mixtures and the multivariate neighbor models, when the latter incorporate a log-linear specification of the correlation structure in the matched pairs.


In scenario 2, the proposed estimators still offer the smallest mean square errors.
 As before, the univariate model offers the best overall performance in terms of bias, variance and mean square error.
 Of the two multivariate models, the one including the interactions performs better as one might expect, with a bias that is about forty times smaller and a mean square error that is about five times smaller.
 Without surprise, the estimator by \citeA{racinskij_et_al_2019} has a worse performance than in the previous scenario, because the conditional independence assumption is violated in this scenario.
 However, this degradation is such that it has a worst performance than the naive estimator for each performance measure.
 An interesting observation is that it also performs much worse than the estimator based on the multivariate neighbor model without interactions, with a relative bias and mean square error that are bigger by more than an order of magnitude, while this latter estimator performs better than the naive estimator for each performance measure.
 A possible explanation is that the multivariate neighbor model implicitly accounts for the all the interactions in the distribution of the unmatched pairs, while the model by \citeA{racinskij_et_al_2019} ignores these interactions.
 This is a further illustration of the difference between classical log-linear mixtures and multivariate neighbor models.
 As before, the estimators by \citeA{ding_fienberg_1994}, and \citeA{diconsiglio_tuoto_2015} have a worse performance than the neighbor models, due to the variance of the estimated linkage accuracy.
 However, in terms of bias and mean square error, they perform better than the naive estimator and that from \citeA{racinskij_et_al_2019}.
 Scenario 3 differs from scenario 2 by adding a third order interaction with coefficient $1/4$. However, this change has a negligible impact on the obtained results and observed trends.


In scenario 4, the first linkage rule has a small false negative rate of about $0.4\%$ (i.e., an imperfect recall), by not linking the pairs that have no exact agreement on the surname, day of birth or month of birth.
 This has a direct impact on the univariate neighbor estimator, which now has the third smallest mean square error, behind the two multivariate neighbor estimators; the one with the interactions offering the best performance.
 Excluding the pairs with no exact agreement further degrades the performance of the log-linear mixture estimator (compared to the scenarios 3 and 4), which still has the largest mean square error and a worst performance than the naive estimator.
 However this change has a limited impact on the performance of the estimators by \citeA{ding_fienberg_1994}, and \citeA{diconsiglio_tuoto_2015}.
 Scenario 5 differs from scenario 4 by adding a third order interaction with coefficient $1/4$. However, this change has a negligible impact on the results.


In summary, the simulation results demonstrate that the proposed estimators may be used to estimate the coverage with a small relative bias and a smaller mean square error than the alternative estimators proposed by \citeA{ding_fienberg_1994},  \citeA{diconsiglio_tuoto_2015}, and \citeA{racinskij_et_al_2019}, when the false negatives are negligible or the true positive probabilities are constrained by a log-linear specification.

%
\begin{table}[htbp]
\centering
\caption{Simulation results. \label{table: simulation results}}
\vspace{12pt}
\resizebox{\textwidth}{!}{
\begin{tabular}{llccc}\hline
Scenario	&	Estimator	&	Relative bias (\%)	&	Variance $\times 10^{-7}$	&	Mean square error $\times 10^{-7}$	\\ \hline
1	&	Naive	&	-5.522	&	12.90	&	24711.31	\\
	&	R	&	-0.034	&	824.62	&	817.32	\\
	&	DF	&	1.618	&	2377.74	&	4475.68	\\
	&	DT	&	1.159	&	2550.13	&	3613.51	\\
	&	UN	&	-0.003	&	8.43	&	8.35	\\
	&	MN with no interactions	&	-0.023	&	28.25	&	28.41	\\
	&	MN with 2nd order interactions	&	-0.119	&	27.06	&	38.25	\\
	&		&		&		&		\\
2 \& 3	&	Naive	&	-4.994	&	15.10	&	20216.56	\\
	&	R	&	-7.784	&	324.65	&	49403.40	\\
	&	DF	&	1.667	&	3381.80	&	5598.43	\\
	&	DT	&	0.961	&	3560.11	&	4272.01	\\
	&	UN	&	-0.004	&	8.31	&	8.24	\\
	&	MN with no interactions	&	-0.423	&	21.05	&	165.44	\\
	&	MN with 2nd order interactions	&	-0.091	&	23.70	&	30.12	\\
	&		&		&		&		\\
4 \& 5	&	Naive	&	-4.292	&	15.46	&	14934.57	\\
	&	R	&	-3.497	&	280414.54	&	287515.73	\\
	&	DF	&	1.760	&	2255.34	&	4740.96	\\
	&	DT	&	1.159	&	2550.13	&	3613.51	\\
	&	UN	&	-0.393	&	9.30	&	134.50	\\
	&	MN with no interactions	&	-0.423	&	21.05	&	165.44	\\
	&	MN with 2nd order interactions	&	-0.091	&	23.70	&	30.12	\\ \hline
\multicolumn{5}{l}{\footnotesize{DF: estimator by \citeA{ding_fienberg_1994}}}\\
\multicolumn{5}{l}{\footnotesize{DF: estimator by \citeA{diconsiglio_tuoto_2015}}}\\
\multicolumn{5}{l}{\footnotesize{MN: estimator based on the multivariate neighbor model}}\\
\multicolumn{5}{l}{\footnotesize{Naive: Lincoln-Petersen estimator that ignores the linkage errors}}\\
\multicolumn{5}{l}{\footnotesize{R : estimator by \citeA{racinskij_et_al_2019}}}\\
\multicolumn{5}{l}{\footnotesize{UN: estimator based on the univariate neighbor model}}
\end{tabular}}
\end{table}


\section{Conclusion}

\noindent A new methodology has been described for capture-recapture estimation with linkage errors, which is based on modeling the number of links from a record, without clerical reviews, including a univariate model and a related multivariate model.
 With the univariate model, the coverage is estimated by linking the records with a sufficiently high recall.
 With the multivariate model, the coverage is estimated by constraining the interactions in the matched pairs through a log-linear specification, while allowing arbitrary interactions in the unmatched pairs; a major difference with classical log-linear mixtures.
 In this latter case, the records must be linked with a high precision to obtain a reliable estimate of the coverage.
 Simulations with public census data demonstrate the good performance of the proposed estimators, when compared to previous solutions.
 Future work will look at obtaining variances and confidence intervals, and at validating the log-linear specification when the multivariate model is used.


\bibliography{rl_biblio}


\begin{appendix}


\section{Extension for undercoverage}\label{appendix: undercoverage}

This appendix aims to extend the results from \citeA{dasylva_goussanou_jjsds_2022} to show that
%
%
\begin{eqnarray*}
 \frac{TP}{TP+FN} &\stackrel{p}{\longrightarrow}&
 P(i \in S_A)^{-1} \overline{p}, \\
 \frac{TP}{TP+FP} &\stackrel{p}{\longrightarrow}&
 \frac{\overline{p}}{\overline{p}+\overline{\lambda}}, \\
 \widehat{\overline{p}} &\stackrel{p}{\longrightarrow}&
 \overline{p},\\
 \frac{\widehat{\overline{p}}}{\widehat{\overline{p}}+
 \widehat{\overline{\lambda}}} &\stackrel{p}{\longrightarrow}& 
 \frac{\overline{p}}{\overline{p}+\overline{\lambda}}.
\end{eqnarray*}
 Therefore, $ P(i \in S_A)^{-1} \widehat{\overline{p}}$ and $\widehat{\overline{p}} /(\widehat{\overline{p}}+\widehat{\overline{\lambda}})$ estimate the recall and precision consistently, in the sense that
%
%
\begin{eqnarray*}
 P(i \in S_A)^{-1}\widehat{\overline{p}} - \frac{TP}{TP+FN}
 &\stackrel{p}{\longrightarrow}& 0, \\
 \frac{\widehat{\overline{p}}}{\widehat{\overline{p}}+
 \widehat{\overline{\lambda}}} - \frac{TP}{TP+FP}
 &\stackrel{p}{\longrightarrow}& 0.
\end{eqnarray*}
 The extension consists in accounting for the undercoverage in $S_A$.
 To proceed, some additional notation is needed. Call $V_j'$ a {\it neighbor} of $V_i$, if unit $j$ is  included in $S_A$ and $V_j'$ is contained in the neighborhood of $V_i$, i.e., ${\cal B}_N(V_i)$.
 The neighbor is called matched if both records are from the same unit.
 Otherwise, it is called unmatched.
 Also, define the following additional notation.
For $i,i'\in S_B$ such that $i \neq i'$, let
%
%
\begin{eqnarray}
 \label{eq: n_i^{(0)}}
 n_i^{(0)} &=&
 \sum_{j \in S_A} I \left ( V_j'\in {\cal B}_N(V_i)\right ),\\
%
%
 \label{eq: n_{iU}^{(0)} and n_{iU}}
 \left ( n_{i\vert U}^{(0)},n_{i\vert U}\right ) &=&
\left ( n_i^{(0)},n_i \right ) -
\left ( I \left ( \{i \in S_A\} \cap
 \left \{ V_i'\in {\cal B}_N(V_i) \right \} \right ),
I(i \in S_A)L_{ii}\right ),\\
%
%
 \label{eq: n_{ii'U}^{(0)}}
 n_{ii'\vert U}^{(0)} &=&
 \sum_{t \in S_A: t \neq i,i'}
 I \left ( V_t'\in {\cal B}_N(V_i) \cap
 {\cal B}_N(V_{i'})\right ),
\end{eqnarray}
 where $n_i^{(0)}$ is the number of neighbors of $V_i$, $n_{i\vert U}^{(0)}$ is the number of unmatched neighbors of this record, $n_{i\vert U}$ is the number of unmatched records, which are linked to the same record, and $n_{ii'\vert U}^{(0)}$ is the number of unmatched neighbors, which are common to $V_i$ and $V_{i'}$.
 Corresponding to $n_i^{(0)}$ and $n_{i\vert U}^{(0)}$, let
 $S_{Ai} =
  \left \{ t\in S_A \ s.t. \
  V_t' \in {\cal B}_N \left ( V_i \right ) \right \}$ denote the subset of units, which are included in $S_A$ and have their record in the neighborhood, and let
 $S_{Ai\vert U} = S_{Ai} - \{i\}$ denote the subsets of these units that are different from unit $i$.
 These latter units are associated with the unmatched neighbors.
Finally, for random variables (or vectors) $X$, $Y$ and $Z$, denote the independence of $X$ and $Y$ by $X \perp\!\!\!\perp Y$, and their conditional independence given $Z$ by $X \perp\!\!\!\perp Y \vert Z$.

The following lemma extends Lemma 2 in \citeA{dasylva_goussanou_jjsds_2022}.
%

\begin{lemma}\label{lemma: condition expectation of Z_i Z_{i'}}

\noindent Suppose that $\left [ \left ( I(i\in S_A), I(i\in S_B), V_i, V_i'\right )\right ]_{1 \leq i \leq N}$ are iid and let $Z_1,\ldots,Z_N$ denote identically distributed random variables, such that they are conditionally independent given $\left [ \left ( I(i \in S_A),I(i \in S_B),V_i,V_i'\right )\right ]_{1\leq i \leq N}$ with a marginal conditional distribution of $Z_i$ that is only a function of  $I(i \in S_A)$, $I(i\in S_B)$, $V_i$, $S_{Ai\vert U}$, $\left [ V_t'\right ]_{t\in S_{Ai \vert U}}$, and $V_i'$. Then
%
%
\begin{equation}
\label{eq: conditional independence 1}
\left ( I(i'\in S_A), V_{i'}',\left [ V_t' \right ]_{t \in S_{Ai'|U}} \right )
\perp\!\!\!\perp
\left ( I(i\in S_A), V_i',\left [ V_t' \right ]_{t \in S_{Ai|U}} \right )
\left \vert 
\begin{array}{l}
i\in S_B,V_i,S_{Ai\vert U},\\
i'\in S_B,V_{i'},S_{Ai'\vert U},\\
{\cal B}_N(V_i)\cap {\cal B}_N(V_{i'})=\varnothing,\\
V_i'\notin {\cal B}_N(V_{i'}),\\
V_{i'}'\notin {\cal B}_N(V_{i})
\end{array}
\right .,
\end{equation}
%
%
\begin{equation}
\label{eq: conditional independence 2}
S_{Ai'\vert U}
\perp\!\!\!\perp
\left ( I(i\in S_A), V_i',\left [ V_t' \right ]_{t \in S_{Ai|U}} \right )
\left \vert 
\begin{array}{l}
i\in S_B,V_i,S_{Ai\vert U},\\
i'\in S_B,V_{i'},\\
{\cal B}_N(V_i)\cap {\cal B}_N(V_{i'})=\varnothing,\\
V_i'\notin {\cal B}_N(V_{i'}),\\
V_{i'}'\notin {\cal B}_N(V_{i})
\end{array}
\right . .
\end{equation}

\noindent Also for any fixed $(v_i,v_{i'}) \in {\cal V}_N^{\ast} \times {\cal V}_N^{\ast}$, $a_i \in \{0,1\}$, $w_i  \notin {\cal B}_N(v_{i'})$, $s_{Ai\vert U}\subset \{1,\ldots,N\}\setminus \{i,i'\}$ and
$\left [ w_t \right ]_{t \in s_{Ai\vert U}}$
 such that
 ${\cal B}_N(v_i) \cap {\cal B}_N(v_{i'})=\varnothing$ and $w_t \in {\cal B}_N(v_i)$ for all $t \in s_{Ai\vert U}$
%
%
\begin{equation}
\label{eq: conditional independence 3}
P \left (
\begin{array}{l}
I(i \in S_A)= a_i,\\
V_i'=w_i,\\
\left [ V_t' \right ]_{t \in s_{Ai \vert U}}=\\
\left [ w_t \right ]_{t \in s_{Ai \vert U}}
\end{array}
\left \vert 
\begin{array}{l}
i\in S_B,V_i=v_i,\\
S_{Ai\vert U}=s_{Ai\vert U},\\
i'\in S_B,V_{i'}=v_{i'}\\
{\cal B}_N(V_i)\cap {\cal B}_N(V_{i'})=\varnothing,\\
V_i'\notin {\cal B}_N(V_{i'}),\\
V_{i'}'\notin {\cal B}_N(V_{i})
\end{array}
\right . \right ) =
P \left ( \left .
\begin{array}{l}
I(i \in S_A)= a_i,\\
V_i'=w_i,\\
\left [ V_t' \right ]_{t \in s_{Ai \vert U}}=\\
\left [ w_t \right ]_{t \in s_{Ai \vert U}}
\end{array}
\right \vert 
\begin{array}{l}
i \in S_B,V_i=v_i,\\
S_{Ai\vert U}=s_{Ai\vert U},\\
V_i'\notin {\cal B}_N(v_{i'})
\end{array}
\right ).
\end{equation}
 Hence
%
%
\begin{equation}
\label{eq: old lemma 3 E[Z_iZ_{i'}]}
E \left [
Z_i Z_{i'}
\left \vert 
\begin{array}{l}
i\in S_B,V_i=v_i, \ 
n_{i\vert U}^{(0)}=k,\\
i'\in S_B,V_{i'}=v_{i'}, \ 
n_{i'\vert U}^{(0)}=\ell,\\
{\cal B}_N(V_i)\cap {\cal B}_N(V_{i'})=\varnothing,\\
V_i' \notin {\cal B}_N(V_{i'}), \ 
V_{i'}' \notin {\cal B}_N(V_{i})
\end{array}
\right . \right ] =
g(v_{i},k;v_{i'})
g(v_{i'},\ell;v_i),
\end{equation}
 where
\begin{equation}
g(v_{i},k;v_{i'})=
E \left [ Z_i
\left \vert 
\begin{array}{l}
i \in S_B,V_i=v_i,\\
n_{i \vert U}^{(0)}=k, \ V_i'\notin {\cal B}_N(v_{i'})
\end{array}
\right . \right ]
\end{equation}

\end{lemma}


\noindent \underline{Proof}:
As in Lemma 2 from \citeA{dasylva_goussanou_jjsds_2022}, recall that for six nonempty events $A_1,A_2$,$B_1,B_2$,$C_1$ and $C_2$ such that $(A_1,A_2)$, $(B_1,B_2)$ and $(C_1,C_2)$ are mutually independent, we have the conditional independence of $A_1$, $B_1$ and $C_1$ given $A_2$, $B_2$ and $C_2$, i.e.,
\begin{eqnarray}
 P \left ( A_1 \cap B_1 \cap C_1 \vert  A_2 \cap B_2 \cap C_2 \right ) &=&
 P(A_1\vert A_2 \cap B_2 \cap C_2 )
 P(B_1\vert A_2 \cap B_2 \cap C_2 )\times \nonumber \\
 & &
 \label{eq: lemma 2 key property}
 P(C_1\vert A_2 \cap B_2 \cap C_2 ).
\end{eqnarray}
 In what follows, consider fixed $a_i, a_{i'} \in \{0,1\}$, $ s_{Ai\vert U}, s_{Ai'\vert U} \subset \{1,\ldots,N\}\setminus \{i,i'\}$, $(v_i,v_{i'}) \in {\cal V}_N^{\ast} \times {\cal V}_N^{\ast}$, $w_i  \notin {\cal B}_N(v_{i'})$, $w_{i'} \notin {\cal B}_N(v_i)$, 
$\left [ w_t \right ]_{t \in s_{Ai\vert U}}$
and $\left [ w_t \right ]_{t \in s_{Ai'\vert U}}$ such that ${\cal B}_N(v_i)\cap {\cal B}_N(v_{i'})=\varnothing$, $s_{Ai\vert U}\cap s_{Ai'\vert U}=\varnothing$, $w_t \in {\cal B}_N(v_i)$ if $t \in s_{Ai\vert U}$ and $w_t \in {\cal B}_N(v_{i'})$ if $t \in s_{Ai'\vert U}$.

\noindent To prove Eq.~\ref{eq: conditional independence 1}
, let $C_1$ denote the certain event (i.e. $P(C_1)=1$) and define the other events as follows.
\begin{eqnarray}
 \label{eq: first A_1}
 A_1 &=&
 \left \{ I(i\in S_A)=a_i  \right \} \cap
 \left \{ V_i'=w_i  \right \} \cap
 \left \{ \left [ V_t'\right ]_{t\in s_{Ai\vert U}}=\left [ w_t\right ]_{t\in s_{Ai\vert U}} \right \}, \\
 \label{eq: first A_2}
 A_2 &=&
 \left \{ i\in S_B, V_i=v_i\right \} \cap
 \{ V_i'\notin {\cal B}_N(v_{i'}) \} \cap
 \left \{ \bigcap_{t\in s_{Ai\vert U}}
 \left \{ t \in S_A \ and \ V_t'\in {\cal B}_N(v_i) \right \}
 \right \}, \\
 B_1 &=&
 \left \{ I(i'\in S_A)=a_{i'}  \right \} \cap
 \left \{ V_{i'}'=w_{i'}  \right \} \cap
 \left \{ \left [ V_t'\right ]_{t\in s_{Ai'\vert U}}=\left [ w_t \right ]_{t\in s_{Ai'\vert U}} \right \}, \nonumber \\
 B_2 &=&
 \left \{ i'\in S_B, V_{i'}=v_{i'}\right \} \cap
 \{ V_{i'}'\notin {\cal B}_N(v_{i}) \} \cap
 \left \{ \bigcap_{t\in s_{Ai'\vert U}}
 \left \{ t \in S_A \ and \ V_t'\in {\cal B}_N(v_{i'}) \right \}
 \right \}, \nonumber \\
 C_2 &=&
 \bigcap_{t\in \{1,\ldots,N\}\setminus \left ( \{i,i'\}\cup s_{Ai\vert U}\cup s_{Ai'\vert U} \right )}
 \left \{ t \notin S_A \ or \ V_t'\notin
 {\cal B}_N(v_i) \cup {\cal B}_N(v_{i'}) \right \}. \nonumber
\end{eqnarray}
 Then, the mutual independence of $(A_1,A_2)$, $(B_1,B_2)$ and $(C_1,C_2)$ follows from that of $\left [ \left ( I(i\in S_A), I(i\in S_B), V_i, V_i'\right )\right ]_{1 \leq i \leq N}$ and Eq.~\ref{eq: conditional independence 1} follows from Eq.~ \ref{eq: lemma 2 key property}.

\noindent For Eq.~\ref{eq: conditional independence 2}, define $A_1$ and $A_2$ as in Eqs.~\ref{eq: first A_1}-\ref{eq: first A_2},
let $B_1$ be the certain event and define the other events as follows.
\begin{eqnarray*}
 B_2 &=&
 \left \{ i'\in S_B, V_{i'}=v_{i'}\right \} \cap
 \{ V_{i'}'\notin {\cal B}_N(v_{i}) \}\\
 C_1 &=&
 \left \{ \bigcap_{t\in s_{Ai'\vert U}}
 \left \{ t \in S_A \ and \ V_t'\in {\cal B}_N(v_{i'}) \right \}
 \right \}\\
 C_2 &=&
 \bigcap_{t\in \{1,\ldots,N\}\setminus \left ( \{i,i'\} \cup s_{Ai\vert U} \right )} \left \{ t \notin S_A \ or \ V_t'\notin {\cal B}_N(v_i) \right \}.
\end{eqnarray*}
 As before, the mutual independence of $(A_1,A_2)$, $(B_1,B_2)$ and $(C_1,C_2)$ follows from that of $\left [ \left ( I(i\in S_A), I(i \in S_B), V_i, V_i'\right )\right ]_{1 \leq i \leq N}$ so that Eq.~\ref{eq: conditional independence 2} follows from Eq.~\ref{eq: lemma 2 key property}.

\noindent Obtaining Eq.~\ref{eq: conditional independence 3} is straightforward  by using the fact that $\left ( I(i'\in S_A), I(i'\in S_B), V_{i'},V_{i'}'\right )$ is independent of $\left ( I(i\in S_A), I(i\in S_B), V_{i},V_i'\right )$ and $\left [  V_t'\right ]_{t \in \{1,\ldots,N\}\setminus \{i,i'\}}$.

\noindent To prove Eq.~\ref{eq: old lemma 3 E[Z_iZ_{i'}]}, let $X$ denote the left-hand side of Eq.~\ref{eq: old lemma 3 E[Z_iZ_{i'}]}. Then
%
%
\begin{eqnarray*}
X =
E \left [
E \left [ Z_i Z_{i'}
\left \vert 
\begin{array}{l}
i\in S_B,V_i,S_{Ai\vert U},\\
i'\in S_B,V_{i'},S_{Ai'\vert U},\\
{\cal B}_N(V_i)\cap {\cal B}_N(V_{i'})=\varnothing,\\
V_i'\notin {\cal B}_N(V_{i'}),\\
V_{i'}'\notin {\cal B}_N(V_{i})
\end{array}
\right . \right ]
\left \vert 
\begin{array}{l}
i\in S_B,V_i=v_i,
n_{i\vert U}^{(0)}=k,\\
i'\in S_B,V_{i'}=v_{i'},
n_{i'\vert U}^{(0)}=\ell,\\
{\cal B}_N(V_i)\cap {\cal B}_N(V_{i'})=\varnothing,\\
V_i' \notin {\cal B}_N(V_{i'}),\\
V_{i'}' \notin {\cal B}_N(V_{i})
\end{array}
\right . \right ]
\end{eqnarray*}
The properties of $Z_i$ and $Z_{i'}$ and Eq.~\ref{eq: conditional independence 1} imply that
%
%
\begin{equation*}
\resizebox{\textwidth}{!}
{$
X =
E \left [
\underbrace{
E \left [ Z_i
\left \vert 
\begin{array}{l}
i\in S_B,V_i,S_{Ai\vert U},\\
i'\in S_B,V_{i'},S_{Ai'\vert U},\\
{\cal B}_N(V_i)\cap {\cal B}_N(V_{i'})=\varnothing,\\
V_i'\notin {\cal B}_N(V_{i'}),\\
V_{i'}'\notin {\cal B}_N(V_{i})
\end{array}
\right . \right ]}_{=(I)} \times \right . 
\left .
\underbrace{
E \left [ Z_{i'}
\left \vert 
\begin{array}{l}
i\in S_B,V_i,S_{Ai\vert U},\\
i'\in S_B,V_{i'},S_{Ai'\vert U},\\
{\cal B}_N(V_i)\cap {\cal B}_N(V_{i'})=\varnothing,\\
V_i'\notin {\cal B}_N(V_{i'}),\\
V_{i'}'\notin {\cal B}_N(V_{i})
\end{array}
\right . \right ]}_{=(II)}
\left \vert 
\begin{array}{l}
i\in S_B,V_i=v_i,
n_{i\vert U}^{(0)}=k,\\
i'\in S_B,V_{i'}=v_{i'},
n_{i'\vert U}^{(0)}=\ell,\\
{\cal B}_N(V_i)\cap {\cal B}_N(V_{i'})=\varnothing,\\
V_i' \notin {\cal B}_N(V_{i'}),\\
V_{i'}' \notin {\cal B}_N(V_{i})
\end{array}
\right . \right ],
$}
\end{equation*}
Using Eqs.~\ref{eq: conditional independence 2}-\ref{eq: conditional independence 3} and the symmetry of all choices of $S_{Ai\vert U}$ such that $n_{i\vert U}^{(0)}=k$, we have
%
%
\begin{equation*}
(I) =
E \left [ Z_i
\left \vert 
\begin{array}{l}
i \in S_B,V_i=v_i,\\
S_{Ai\vert U}, \ V_i'\notin {\cal B}_N(v_{i'})
\end{array}
\right . \right ]=
\underbrace{
E \left [ Z_i
\left \vert 
\begin{array}{l}
i \in S_B,V_i=v_i,\\
n_{i \vert U}^{(0)}=k, \ V_i'\notin {\cal B}_N(v_{i'})
\end{array}
\right . \right ]}_{=g(v_i,k;v_{i'})}.
\end{equation*}
 In a similar manner $(II)=g(v_{i'},\ell;v_i)$. Therefore $X = g(v_{i},k;v_{i'}) g(v_{i'},\ell;v_i)$.

\hfill QED

The above lemma leads to the following theorem, which extends Theorem 1 in \citeA{dasylva_goussanou_jjsds_2022}.
%
%
\begin{theorem}
\label{theorem: law of large number}

\noindent Consider ${\cal V}_N$, ${\cal B}_N(.)$, $S_A$, $S_B$, and $\left [ Z_i\right ]_{1\leq i \leq N}$ identically distributed random variables such that the following conditions apply.

\begin{enumerate}

\item[(C.1)] $\lim_{N\rightarrow \infty}
 P \left ( i\in S_B\right ) = \tau$ for some positive $\tau$.

\item[(C.2)] For $\Lambda \geq 0$ not depending on $N$,
 $\sup_{v \in {\cal V}_N^{\ast}} (N-1) \lambda_N^{(0)} \left ( v \right ) \leq \Lambda$.

\item[(C.3)] For $c \geq 0$ not depending on $N$
%
%
\begin{eqnarray*}
%
%
 N P \left ( {\cal B}_N \left ( V_i \right ) \cap 
 {\cal B}_N \left ( V_{i'} \right ) \neq \varnothing
 \big \vert \{i,i'\} \subset S_B
 \right ) &\leq& c, \\
%
%
%
N P \left (
 V_i' \in {\cal B}_N(V_{i'}) \left \vert
 \{i,i'\} \subset S_B,
 {\cal B}_N(V_i) \cap {\cal B}_N(V_{i'})=\varnothing
 \right . \right ) &\leq& c.
\end{eqnarray*}

\item[(C.4)] $Z_1,\ldots,Z_N$ are conditionally independent given $\left [ \left ( I(i \in S_A),I(i \in S_B),V_i,V_i'\right )\right ]_{1\leq i \leq N}$ such that the marginal conditional distribution of $Z_i$ is only a function of  $I(i \in S_A)$, $I(i\in S_B)$, $V_i$, $S_{Ai\vert U}$, $\left [ V_t'\right ]_{t\in S_{Ai \vert U}}$, and $V_i'$.

\item[(C.5)] $\vert Z_i\vert \leq R_N \left ( n_{i\vert U}^{(0)} \right )$ where $R_N(.)$ is a polynomial with a finite degree not depending on $N$ and nonnegative coefficients of $O\left ( \left ( \log N\right )^d \right )$, where $d$ does not depend on $N$ either.

\item[(C.6)]
$ \lim_{N \rightarrow \infty} E \left [ Z_i \left \vert i \in S_B \right . \right ] = \mu$, where $\vert\mu\vert<\infty$.

\end{enumerate}

\noindent Then
%
%
\begin{equation}
\label{eq: general convergence}
 \frac{1}{\vert S_B \vert}
 \sum_{i \in S_B} Z_i \stackrel{p}{\longrightarrow} \mu.
\end{equation}

\end{theorem}


\noindent \underline{Proof}: 
 The proof is identical to that of Theorem 1 in \citeA{dasylva_goussanou_jjsds_2022}, after aligning the notation (including setting $\Pi(.)$ to the identity and replacing $\phi_N$ by $S_B$) and replacing Lemma 2 (from \citeA{dasylva_goussanou_jjsds_2022}) by Lemma~\ref{lemma: condition expectation of Z_i Z_{i'}}.
 Also, ${\cal B}_N(v)$, $p_N(v)$, $\lambda_N(v)$, $\lambda_N^{(0)}(v)$,$n_i^{(0)}$, $\left (n_{i\vert U}^{(0)}, n_{i\vert U}\right )$ and $n_{ii'\vert U}^{(0)}$ are now given by Eqs.~\ref{eq: B_N(v)}, \ref{eq: p_N(v)}, \ref{eq: lambda_N(v)}, \ref{eq: lambda_N^{(0)}(v)}, \ref{eq: n_i^{(0)}}, \ref{eq: n_{iU}^{(0)} and n_{iU}} and \ref{eq: n_{ii'U}^{(0)}}, respectively.

\hfill QED

The next result shows the convergence of the recall and precision to $P(i \in S_A)^{-1} \overline{p}$ and $\overline{p}/(\overline{p}+\overline{\lambda})$, respectively.
 It extends Corollary 1 in \citeA{dasylva_goussanou_jjsds_2022} and is a direct consequence of Theorem~\ref{theorem: law of large number}.
%
%
\begin{corollary}\label{corollary: lln error rates}

\noindent Suppose that assumptions C.1-C.3 hold and that the linkage meets the following conditions
\begin{enumerate}

\item[(C.7)] $\left [ L_{1j} \right ]_{1\leq j \leq N}$,...,$\left [ L_{N j} \right ]_{1\leq j \leq N}$  are conditionally independent given

\noindent
 $\left [ \left ( I(i \in S_A), I(i \in S_B), V_i,
 V_{\Pi(i)}'\right ) \right ]_{1\leq i \leq N}$,
 where the conditional distribution of
 $\left [ L_{ij} \right ]_{1\leq j \leq N}$ is only a function of
 $I(i \in S_A)$, $I(i \in S_B)$, $V_i$, $S_{Ai \vert U}$,
 $\left [ V_t'\right ]_{t \in S_{Ai \vert U}}$, and $V_i'$.

\item[(C.8)]
\begin{equation}
\lim_{N \rightarrow \infty} E \left [ \left . \left ( p_N(V_i), (N-1)\lambda_N(V_i) \right )\right \vert
 i\in S_B \right ] = \left ( \overline{p},\overline{\lambda}\right ).
\end{equation}
\end{enumerate}

\noindent Then
%
%
\begin{equation}
\label{eq: lln for the recall and precision}
 \left ( \frac{TP}{TP+FN},\frac{TP}{TP+FP}\right )
 \stackrel{p}{\longrightarrow}
 \left ( \frac{\overline{p}}{P(i \in S_A)},
 \frac{\overline{p}}{\overline{p}+\overline{\lambda}}\right ).
\end{equation}
 In particular, Eq.~\ref{eq: lln for the recall and precision} holds if C.8 is replaced by the condition
$$\left . \left ( p_N(V_i),(N-1)\lambda_N(V_i)\right ) \right \vert
  \left \{ i\in S_B \right \} \stackrel{d}{\longrightarrow} F(.,.)$$
 with $\overline{p}=\int p dF(p,\lambda)$ and
 $\overline{\lambda}=\int \lambda dF(p,\lambda)$.

\end{corollary}


\noindent \underline{Proof}:
 The proof is similar to that of Corollary 1 in \citeA{dasylva_goussanou_jjsds_2022}.
 Indeed, the only change to the proof is regarding the convergence of $TP/(TP+FN)$.
 In that regard, note that
 $TP+FN=|S_A \cap S_B|$ (i.e., the total number of matched pairs) and
%
%
\begin{equation*}
 \frac{TP}{TP+FN} =
 \frac{TP}{|S_B|} \left ( \frac{|S_A \cap S_B|}{|S_B|}\right )^{-1},
\end{equation*}
 where $|S_A \cap S_B|/|S_B|$ and $TP/|S_B|$ converge in probability to $P(i \in S_A \left | i \in S_B \right . ) = P(i \in S_A)$ and $\overline{p}$, respectively; the latter convergence resulting from an application of Theorem~\ref{theorem: law of large number} with $Z_i=I(i \in S_A) L_{ii}$.

\hfill QED

Other results from \citeA{dasylva_goussanou_jjsds_2022} remain valid, based on Theorem~\ref{theorem: law of large number} and Corollary~\ref{corollary: lln error rates}, such as Lemma 2 (convergence in distribution of $n_i$ to a mixture as in the right-hand side of Eq.~\ref{eq: limiting mixture}), Theorem 2 (consistency of the estimator by \citeA{blakely_salmond_2002} in the homogeneous case) and Theorem 3 (consistency of the maximum likelihood estimator).
 This is easily seen by inspecting the related proofs in \citeA{dasylva_goussanou_jjsds_2022}.
 Therefore, the estimators $\widehat{\overline{p}}$ and $\widehat{\overline{\lambda}}$ are consistent, and $P(i \in S_A)^{-1} \widehat{\overline{p}}$ and $\widehat{\overline{p}}/(\widehat{\overline{p}}+\widehat{\overline{\lambda}})$ are consistent estimators of the recall and precision, respectively.


\section{Multivariate extension}\label{appendix: multivariate}

To describe the multivariate version of the neighbor model, let $\Gamma$ denote the index set\footnote{To avoid any conflict with the previously defined notation, it is assumed that the rules are indexed such that $\Gamma$ does not contain $0$.} of a collection of rules and let $L_{ij}^{(\gamma)}$ indicate whether the pair $(i,j)$ is linked by  rule $\gamma \in \Gamma$.
 The set $\Gamma$ may take various forms, such as a subset of consecutive integers starting from $1$, or it can be a subset of $\{0,1\}^K$ if linking the records with $K$ linkage variables and performing an exact comparison for each variable.
 For $\gamma \in \Gamma$, let $n_i^{(\gamma)} = \sum_{j \in S_A} L_{ij}^{(\gamma)}$, $\bm{n}_i = \left [ n_i^{(\gamma)}\right ]_{\gamma \in \Gamma}$, and redefine ${\cal B}_N(v)$ to be a subset of ${\cal V}_N$, which satisfies the condition
%
%
\begin{equation*}
 \label{eq: multivariate B_N(v)}
 {\cal B}_N(v) \supset
 \left \{ v' \in {\cal V}_N \ s.t. \
 E \left [ \left . \sum_{\gamma \in \Gamma} L_{ij}^{(\gamma)}
 \right | (i,j) \in S_B \times S_A, \left ( V_i,V_j'\right )=(v,v')
 \right ]>0 \right \}.
\end{equation*}
 In words, ${\cal B}_N(v)$ is a superset of record values, which are linked by at least one rule in the collection with a positive probability, given that $i \in S_B$ and $V_i=v$.
 The function $\lambda_N^{(0)}(.)$ is still defined by Eq.~\ref{eq: lambda_N^{(0)}(v)}, while $p_N(.)$ and $\lambda_N(.)$ are replaced by the vectors $\bm{p}_N(v)=\left [ p_N^{(\gamma)}(v)\right ]_{\gamma \in \Gamma}$ and $\bm{\lambda}^{(\gamma)}_N(v)=\left [ \lambda_N^{(\gamma)}(v)\right ]_{\gamma \in \Gamma}$, with
%
%
\begin{eqnarray}
 \label{eq: multivariate p_N(v)}
 p^{(\gamma)}_N(v) &=&
 E \left [ \left . I \left ( i \in S_A \right ) L_{ii}^{(\gamma)}
 \right | i \in S_B, V_i=v \right ],\\
%
%
 \label{eq: multivariate lambda_N(v)}
 \lambda^{(\gamma)}_N(v) &=&
 E \left [ \left . I \left (j \in S_A \right ) L_{ij}^{(\gamma)}
 \right | i \in S_B, V_i=v \right ], \ j \neq i,
\end{eqnarray}
 and $\sum_{\gamma \in \Gamma} p_N^{(\gamma)}(v) \leq 1$ for all $v \in {\cal V}_N^{\ast}$.
 In words, $p^{(\gamma)}_N(v)$ and $\lambda^{(\gamma)}_N(v)$ are the expected numbers of true positives and false positives for rule $\gamma$, given that $i \in S_B$ and $V_i=v$.
 The regularity condition of Eq.~\ref{eq: census regularity condition 2} is replaced by the following more general condition.
%
%
\begin{eqnarray}
 \label{eq: multivariate census regularity condition 2}
 \left . \left ( \bm{p}_N(V_i),
 (N-1)\bm{\lambda}_N(V_i)\right ) \right |
 \{i \in S_B \} &\stackrel{d}{\longrightarrow}& F.
\end{eqnarray}
 A case of special interest  is when $\bm{p}_N(v)$ is of the form
 $\varrho \left ( \bm{\beta}_N(v)\right )$, for some function
 $\bm{\beta}_N: {\cal V}^{\ast} \rightarrow \mathbb{R}^m$, and some {\it injective} function $\varrho: \mathbb{R}^m \rightarrow [0,1]^{|\Gamma|}$ independent of $N$, where $m<|\Gamma|$.
 In this case, Eq.~\ref{eq: census regularity condition 2} is replaced by the following condition instead.
%
%
\begin{eqnarray}
 \label{eq: multivariate census regularity condition 2 parametric p}
 \left . \left ( \bm{\beta}_N(V_i),
 (N-1)\bm{\lambda}_N(V_i)\right ) \right |
 \{i \in S_B \} &\stackrel{d}{\longrightarrow}& H,
\end{eqnarray}
 where $H$ does not depend on $N$.

The next lemma states the convergence of $\bm{n}_i$ to a multivariate mixture, when $N \rightarrow \infty$ under the conditions given by Eq.~\ref{eq: census regularity condition 1}, and Eq.~\ref{eq: multivariate census regularity condition 2} or Eq.~\ref{eq: multivariate census regularity condition 2 parametric p}.
 The mixing distribution is given by $F$ or $H$ depending on whether Eq.~\ref{eq: multivariate census regularity condition 2} or Eq.~\ref{eq: multivariate census regularity condition 2 parametric p} applies.
 In both cases, the component distributions come from $|\Gamma|$-variate families of discrete distributions, which correspond to the convolution of a multinomial distribution with a product of independent Poisson distributions.
 To further describe the limiting distributions, let ${\cal F}$ denote the family of the component distributions based on Eq.~\ref{eq: multivariate census regularity condition 2}, where each member is of the form
\[ IMultinomial(1,\bm{p}) \ast PPoisson(\bm{\lambda}),\]
 for $\bm{p}=\left [ p^{(\gamma)} \right ]_{\gamma \in \Gamma}$ and $\bm{\lambda}= \left [ \lambda^{(\gamma)} \right ]_{\gamma \in \Gamma}$.
 When Eq.~\ref{eq: multivariate census regularity condition 2 parametric p} applies, the component distributions come from the subset ${\cal F}_{\varrho}$ of ${\cal F}$, where $\bm{p}=\varrho(\bm{\beta})$ for some $\bm{\beta} \in \mathbb{R}^m$.
 A member of this family is a parametric distribution with parameters
 $\bm{\beta} \in \mathbb{R}^m$ and
 $\bm{\lambda} \in (0,+\infty)^{|\Gamma|}$.


\begin{lemma}\label{lemma: multivariate convergence in distribution}

Suppose that Eq.~\ref{eq: census regularity condition 1} applies.
 If Eq.~\ref{eq: multivariate census regularity condition 2} also applies, $\bm{n}_i$ converges in distribution to the mixture of distributions from ${\cal F}$, with mixing distribution $F$.
 If Eq.~\ref{eq: multivariate census regularity condition 2 parametric p} also applies, $\bm{n}_i$ instead converges in distribution to the mixture of distributions from
 ${\cal F}_{\varrho}$, with mixing distribution $H$.

\end{lemma}


\noindent \underline{Proof}: In what follows, $\bm{\omega} =
 \left [ \omega^{(\gamma)}\right ]_{\gamma \in \Gamma} \in
 \mathbb{R}^{|\Gamma|}$, $\bm{p} = \left [ p^{(\gamma)}\right]_{\gamma \in \Gamma} \in [0,1]^{|\Gamma|}$ and $\bm{\lambda} = \left [ \lambda^{(\gamma)}\right]_{\gamma \in \Gamma} \in (0,\infty)^{|\Gamma|}$.
 The proof is based on characteristic functions.
 For the first convergence, first note that if $\bm{Z}$ is distributed according to the member of ${\cal F}$ with parameters
 $\bm{p}=\left [ p^{(\gamma)} \right ]_{\gamma \in \Gamma}$ and
 $\bm{\lambda}= \left [ \lambda^{(\gamma)} \right ]_{\gamma \in \Gamma}$, then the associated characteristic function is
%
%
\begin{equation}
 E \left [ e^{\bm{\omega}^{\top} \bm{Z}}\right ] =
 \left ( 1+\sum_{\gamma \in \Gamma}
 p^{(\gamma)} \left ( e^{\jmath \omega^{(\gamma)}} -1 \right )
 \right )
 \prod_{\gamma \in \Gamma}
 \exp \left ( \lambda^{(\gamma)}
 \left ( e^{\jmath \omega^{(\gamma)}}-1\right ) \right ), \
 \bm{\omega} =
 \left [ \omega^{(\gamma)}\right ]_{\gamma \in \Gamma} \in
 \mathbb{R}^{|\Gamma|}.
\end{equation}
 Next, let $\bm{W}$ denote a random vector that is distributed according to the mixture of distributions from ${\cal F}$ with mixing distribution $F$ and observe that its characteristic function is of the following form.
%
%
\begin{equation*}
 E \left [  e^{\jmath \bm{\omega}^{\top} \bm{W}} \right ] =
 {\displaystyle \int}
 \left ( 1 + \sum_{\gamma \in \Gamma}
 p^{(\gamma)} \left ( e^{\jmath \omega^{(\gamma)}} -1\right )
 \right )
 \prod_{\gamma \in \Gamma}
 \exp \left ( \lambda^{(\gamma)}
 \left ( e^{\jmath \omega^{(\gamma)}}-1\right ) \right )
 dF(\bm{p},\bm{\lambda}).
\end{equation*}
 Next, note that
%
%
\begin{equation}
 \bm{n}_i =
 \left [ I(i \in S_A)L_{ii}^{(\gamma)}\right ]_{\gamma \in \Gamma}+
 \sum_{j\in \{1,\ldots,N\}\setminus \{i\}}
 \left [ I(j \in S_A)L_{ij}^{(\gamma)}\right ]_{\gamma \in \Gamma},
\end{equation}
 where the two terms on the right-hand side have independent multinomial distributions conditional on the event $\{ i \in S_B \}$ and $V_i$.
 Thus,
%
%
\begin{equation*}
\resizebox{\textwidth}{!}
{$
 E \left [  \left . e^{\jmath \bm{\omega}^{\top} \bm{n}_i} \right |
 i \in S_B, V_i \right ] =
 \left ( 1 +
 \sum_{\gamma \in \Gamma} p_N^{(\gamma)} (V_i)
 \left ( e^{\jmath \omega^{(\gamma)}} -1 \right ) \right )
 \left ( 1 +
 \sum_{\gamma \in \Gamma} \lambda_N^{(\gamma)} (V_i)
 \left ( e^{\jmath \omega^{(\gamma)}} -1\right ) \right )^{N-1}.
$}
\end{equation*}
 If $F_N$ denote the conditional distribution of $\left ( \bm{p}_N(V_i),(N-1) \bm{\lambda}_N(V_i)\right )$ given that $i \in S_B$, it follows that
%
%
\begin{equation*}
\resizebox{\textwidth}{!}
{$
 E \left [  \left . e^{\jmath \bm{\omega}^{\top} \bm{n}_i} \right |
 i \in S_B \right ] =
 {\displaystyle \int}
 \left ( 1 +
 \sum_{\gamma \in \Gamma} p^{(\gamma)}
 \left ( e^{\jmath \omega^{(\gamma)}} - 1 \right )\right )
 \left ( 1 +
 \sum_{\gamma \in \Gamma} \frac{\lambda^{(\gamma)}}{N-1}
 \left ( e^{\jmath \omega^{(\gamma)}} -1\right )\right )^{N-1}
 dF_N(\bm{p},\bm{\lambda}).
$}
\end{equation*}
 For notational convenience, let $\Delta \left ( \bm{\omega} \right )=E \left [  \left . e^{\jmath \bm{\omega}^{\top} \bm{n}_i} \right | i \in S_B \right ]-E \left [  e^{\jmath \bm{\omega}^{\top} \bm{W}} \right ]$.
 Then
%
%
\begin{equation*}
\resizebox{\textwidth}{!}
{$
\begin{split}
 \left | \Delta \left ( \bm{\omega} \right ) \right |
 \quad \leq& \quad
 \underbrace{\left | {\displaystyle \int}
 \left ( 1 +
 \sum_{\gamma \in \Gamma} p^{(\gamma)}
 \left ( e^{\jmath \omega^{(\gamma)}} - 1\right ) \right )
 \prod_{\gamma \in \Gamma}
 \exp \left ( \lambda^{(\gamma)}
 \left ( e^{\jmath \omega^{(\gamma)}}-1\right ) \right )
 \left ( dF_N(\bm{p},\bm{\lambda}) -
 dF(\bm{p},\bm{\lambda}) \right ) \right |}_{(I)} + \\
 & \qquad
 \underbrace{{\displaystyle \int}
 \left | \left (1+
 \sum_{\gamma \in \Gamma} \frac{\lambda^{(\gamma)}}{N-1}
 \left ( e^{\jmath \omega^{(\gamma)}} -1 \right ) \right )^{N-1} -
 \prod_{\gamma \in \Gamma} \exp \left ( \lambda^{(\gamma)}
 \left ( e^{\jmath \omega^{(\gamma)}}-1\right ) \right )
 \right | dF_N(\bm{p},\bm{\lambda})}_{(II)},
\end{split}
$}
\end{equation*}
 where $(I) \rightarrow 0$ because $F_N$ converges to $F$ according to Eq.~\ref{eq: multivariate census regularity condition 2}, and $(II) = O(1/N)$ due to Eq.~\ref{eq: census regularity condition 1}.
 Thus $\left | \Delta \left ( \bm{\omega} \right ) \right | \rightarrow 0$, which implies the desired result by the Continuity Theorem \cite[p. 383]{billingsley_1995}. The proof of the second convergence is similar.

\hfill QED

\vspace{6pt}

The next lemma extends Lemma 4 from \citeA{dasylva_goussanou_jjsds_2022}. It gives sufficient conditions for the identification of finite mixtures over ${\cal F}$ (or ${\cal F}_{\varrho}$); a key property for proving the consistency of maximum likelihood estimators.
 The lemma requires a lexicographic order over $(0,\infty)^{|\Gamma|}$.
 To do so, order the elements of $\Gamma$ based on some bijection from $\{1,\ldots,|\Gamma|\}$ into $\Gamma$, which is denoted by $\gamma(.)$ with a slight abuse of the notation.
 Next, denote a tuple $\left [ \lambda^{(\gamma)} \right ]_{\gamma \in \Gamma}$ equivalently by $\left [ \lambda^{(\gamma(t))} \right ]_{1 \leq t \leq |\Gamma|}$, and write $\bm{\lambda} \succ \bm{\lambda}'$ (i.e., $\bm{\lambda}$ greater than $\bm{\lambda}'$), if $\lambda^{(\gamma(1))}>\lambda^{'(\gamma(1))}$ or if there exists $t_0=2,\ldots, |\Gamma|$ such that $\lambda^{(\gamma(t))}=\lambda^{'(\gamma(t))}$ for $t<t_0$ and $\lambda^{(\gamma(t_0))}>\lambda^{'(\gamma(t_0))}$.
 Also, let
 $\max(\bm{\lambda},\bm{\lambda}')=\bm{\lambda}$ if
 $\bm{\lambda} \succ \bm{\lambda}'$ otherwise let
 $\max(\bm{\lambda},\bm{\lambda}')=\bm{\lambda}'$.


\begin{lemma}\label{lemma: identification multivariate}

For positive integers $G$ and $G'$, let
 $\bm{\lambda}_1 \succ \ldots \succ \bm{\lambda}_G$, and
 $\bm{\lambda}_1' \succ \ldots \succ \bm{\lambda}_G'$, and denote by
 $h_g$ and $h_g'$ the members of ${\cal F}$ with the parameters
 $\left ( \bm{p}_g, \bm{\lambda}_g \right )$ and
 $\left ( \bm{p}_g', \bm{\lambda}_g' \right )$, respectively, and suppose that the mixtures $h=\sum_{g=1}^G \alpha_g h_g$ and $h'=\sum_{g=1}^{G'} \alpha_g' h_g'$ are equal, where $\alpha_g$ and $\alpha_g'$ are positive for each $g$.
 Then $G=G'$, $\alpha_g=\alpha_g'$ and
$\left ( \bm{p}_g, \bm{\lambda}_g \right )=
 \left ( \bm{p}_g', \bm{\lambda}_g'\right )$, for $g=1,\ldots,G$.
 Furthermore, if there exists an injective function $\varrho: \mathbb{R}^m \rightarrow [0,1]^{|\Gamma|}$ such that $\bm{p}_g=\varrho(\bm{\beta}_g)$ and $\bm{p}_g'=\varrho(\bm{\beta}_g')$ for each $g$, we also have $\bm{\beta}_g=\bm{\beta}_g'$ for each $g$.

\end{lemma}


\noindent \underline{Proof}:
 Since $\varrho(.)$ is assumed to be injective, it easily follows that $\bm{\beta}_g=\bm{\beta}_g'$ for each $g$, if $G=G'$,
 $\alpha_g=\alpha_g'$,
 $\left ( \bm{p}_g, \bm{\lambda}_g \right )=
  \left ( \bm{p}_g', \bm{\lambda}_g'\right )$,
 $\bm{p}_g=\varrho(\bm{\beta}_g)$ and $\bm{p}_g'=\varrho(\bm{\beta}_g')$, for each $g$.

\noindent The next steps adapt the proof of Theorem 2 from \citeA{teicher_1963}, according to whom it is
sufficient to show the following result to complete the proof.
%
%
\begin{equation}
 \label{eq: identification multivariate sufficient result}
 \left ( \alpha_g, \bm{p}_g, \bm{\lambda}_g \right ) =
 \left ( \alpha_g', \bm{p}_g', \bm{\lambda}_g' \right ),
\end{equation}
 for each $g=1,\ldots,\min(G,G')$.
 Indeed, the sufficiency is obvious if $G=G'$.
 Besides, we cannot have $G<G'$, as this would imply that $\sum_{g=G+1}^{G'} \alpha_g' h_g' = 0$, which is impossible because
 $\alpha_g'>0$.
 For a similar reason, we cannot have $G'<G$.

The proof of Eq.~\ref{eq: identification multivariate sufficient result} is done by induction and based on the linear transformation that maps a distribution $f$ over $\mathbb{N}^{|\Gamma|}$ to the following function
%
%
\begin{equation}\label{eq: transform}
 z \mapsto
 \sum_{\bm{n}=\left [ n^{\gamma(t)}
 \right ]_{1 \leq t \leq |\Gamma|} \in \mathbb{N}^{|\Gamma|}}
 f \left ( \bm{n}\right )
 \exp \left ( z\sum_{t=1}^{|\Gamma|}
 (|\Gamma|-t+1) n^{(\gamma(t))} \right ),
\end{equation}
 where the right-hand side is finite, i.e., in an interval of the form $(-\infty,b)$ for some nonnegative $b$ that is possibly equal to $+\infty$.
 This transform is closely related to the moment generating function of the random vector $\bm{n}$.
 Let $\phi_g$, $\phi_g'$, $\phi$ and $\phi'$ denote the transforms  of $h_g$, $h_g'$, $h$ and $h'$,respectively, and observe that these transforms are well defined over $\mathbb{R}$ (i.e., the right-hand side of Eq.~\ref{eq: transform} is finite for each $z \in \mathbb{R}$).
 Indeed
%
%
\begin{eqnarray*}
 \phi_g(z) &=&
 \left ( 1+\sum_{t=1}^{|\Gamma|}
 p_g^{(\gamma(t))} \left ( e^{(|\Gamma|-t+1) z} -1 \right )
 \right ) \xi(z;\bm{\lambda}_g),\\
 \xi(z;\bm{\lambda}_g) &=&
 \exp \left ( \sum_{t=1}^{|\Gamma|}
 \lambda_g^{(\gamma(t))} \left ( e^{(|\Gamma|-t+1) z}-1\right )
 \right ),
\end{eqnarray*}
 with a similar expression for $\phi_g'$.
 Also, by linearity, $\phi = \sum_{g=1}^G \alpha_g \phi_g$, $\phi' = \sum_{g=1}^{G'} \alpha_g' \phi_g'$.

For the first step of the induction proof consider $g=1$ and proceed as follows.
 Since $h=h'$, we have $\phi=\phi'$ and the ordering of  $\bm{\lambda}_1, \ldots, \bm{\lambda}_G$, and $\bm{\lambda}_1', \ldots, \bm{\lambda}_G'$ implies that
%
%
\begin{equation}
\label{eq: proof multivariate identification g=1 first eq}
\resizebox{\textwidth}{!}
{$
\begin{split}
 \alpha_1 \left ( 1+\sum_{t=1}^{|\Gamma|}
 p_1^{(\gamma(t))} \left ( e^{(|\Gamma|-t+1) z} -1 \right ) \right )
 \frac{\xi(z;\bm{\lambda}_1)}{
 \xi(z;\max \left ( \bm{\lambda}_1, \bm{\lambda}_1')\right )}
 (1+o(1)) \quad =& \quad
 \alpha_1' \left ( 1+\sum_{t=1}^{|\Gamma|}
 p_1^{'(\gamma(t))} \left ( e^{(|\Gamma|-t+1) z} -1 \right ) \right )
 \times \\
 & \qquad
 \frac{\xi(z;\bm{\lambda}_1')}{
 \xi(z;\max \left ( \bm{\lambda}_1, \bm{\lambda}_1')\right )}
 (1+o(1)),
\end{split}
$}
\end{equation}
 when $z \rightarrow +\infty$.
 Eq.~\ref{eq: proof multivariate identification g=1 first eq}
 is incompatible with the fact that $\bm{\lambda}_1 \succ \bm{\lambda}_1'$, because its left-hand side would converge to $\alpha_1>0$ (if each $p_1^{(\gamma)}$ is null) or go to $+\infty$ (if some $p_1^{(\gamma)}$ is positive), while its right-hand side would always converge to $0$.
 For similar reasons, we cannot have $\bm{\lambda}_1' \succ \bm{\lambda}_1$. Thus $\bm{\lambda}_1 = \bm{\lambda}_1'$.
 Since $\phi=\phi'$, this equality implies that 
%
%
\begin{equation}
\label{eq: proof multivariate identification g=1 second eq}
 \alpha_1 \left ( 1+\sum_{t=1}^{|\Gamma|}
 p_1^{(\gamma(t))} \left ( e^{(|\Gamma|-t+1) z} -1 \right )
 \right ) + o(1) =
 \alpha_1' \left ( 1+\sum_{t=1}^{|\Gamma|}
 p_1^{'(\gamma(t))} \left ( e^{(|\Gamma|-t+1) z} -1 \right )
 \right ) + o(1),
\end{equation}
 when $z \rightarrow +\infty$. Now suppose that $\alpha_1 p_1^{(\gamma(t))} \neq \alpha_1' p_1^{'(\gamma(t))}$ for some $t$ and let $t_0$ be the smallest such value of $t$.
 Then, Eq.~\ref{eq: proof multivariate identification g=1 second eq}
 implies that
%
%
\begin{equation*}
\resizebox{\textwidth}{!}
{$
\begin{split}
 \alpha_1 \left ( 1+\sum_{t=t_0}^{|\Gamma|}
 p_1^{(\gamma(t))} \left ( e^{(|\Gamma|-t+1) z} -1 \right )
 \right ) e^{-(|\Gamma|-t_0+1) z} + o(1)  \quad =& \quad
 \alpha_1' \left ( 1+\sum_{t=t_0}^{|\Gamma|}
 p_1^{'(\gamma(t))} \left ( e^{(|\Gamma|-t+1) z} -1
 \right ) \right ) \times \\
 & \qquad e^{-(|\Gamma|-t_0+1) z} + o(1),
\end{split}
$}
\end{equation*}
 where the left-hand side goes to $\alpha_1 p_1^{(\gamma(t_0))}$ and  the right-hand side goes to $\alpha_1' p_1^{'(\gamma(t_0))}$ and the two limits must coincide, when $z \rightarrow +\infty$.
 Since this contradicts the assumption about $t_0$, we necessarily have $\alpha_1 p_1^{(\gamma)} = \alpha_1' p_1^{'(\gamma)}$ for each $\gamma$, which implies that $\alpha_1=\alpha_1'$ (based on Eq.~\ref{eq: proof multivariate identification g=1 second eq}) and $p_1^{(\gamma)} = p_1^{'(\gamma)}$ for each $\gamma$.

For the general induction step, consider $g>1$ and suppose that Eq.~\ref{eq: identification multivariate sufficient result} applies for each $g'<g$.
 Hence,
 $\sum_{g'=g}^G \alpha_{g'} h_{g'}=
  \sum_{g'=g}^G \alpha_{g'}' h_{g'}'$ and
 $\sum_{g'=g}^G \alpha_{g'} \phi_{g'}=
  \sum_{g'=g}^G \alpha_{g'}' \phi_{g'}'$.
 Using the ordering of  $\bm{\lambda}_1, \ldots, \bm{\lambda}_G$, and $\bm{\lambda}_1', \ldots, \bm{\lambda}_G'$ as before, we obtain
%
%
\begin{equation}
\label{eq: proof multivariate identification g>1 first eq}
\resizebox{\textwidth}{!}
{$
\begin{split}
 \alpha_g \left ( 1+\sum_{t=1}^{|\Gamma|}
 p_g^{(\gamma(t))} \left ( e^{(|\Gamma|-t+1) z} -1 \right ) \right )
 \frac{\xi(z;\bm{\lambda}_g)}{
 \xi(z;\max \left ( \bm{\lambda}_g, \bm{\lambda}_g')\right )}
 (1+o(1)) \quad =& \quad
 \alpha_g' \left ( 1+\sum_{t=1}^{|\Gamma|}
 p_g^{'(\gamma(t))} \left ( e^{(|\Gamma|-t+1) z} -1 \right ) \right )
 \times \\
 & \qquad
 \frac{\xi(z;\bm{\lambda}_g')}{
 \xi(z;\max \left ( \bm{\lambda}_g, \bm{\lambda}_g')\right )}
 (1+o(1)),
\end{split}
$}
\end{equation}
 when $z \rightarrow +\infty$.
 As before, Eq.~\ref{eq: proof multivariate identification g>1 first eq} is incompatible with the fact that $\bm{\lambda}_g \succ \bm{\lambda}_g'$, because its left-hand side would converge to $\alpha_g>0$ or go to $+\infty$, while its right-hand side would always converge to $0$.
 Likewise, we cannot have $\bm{\lambda}_g' \succ \bm{\lambda}_g$. Hence $\bm{\lambda}_g = \bm{\lambda}_g'$.
 Since
 $\sum_{g'=g}^G \alpha_{g'} \phi_{g'}=
  \sum_{g'=g}^G \alpha_{g'}' \phi_{g'}'$,
 this equality implies that 
%
%
\begin{equation}
\label{eq: proof multivariate identification g>1 second eq}
 \alpha_g \left ( 1+\sum_{t=1}^{|\Gamma|}
 p_g^{(\gamma(t))} \left ( e^{(|\Gamma|-t+1) z} -1 \right )
 \right ) + o(1) =
 \alpha_g' \left ( 1+\sum_{t=1}^{|\Gamma|}
 p_g^{'(\gamma(t))} \left ( e^{(|\Gamma|-t+1) z} -1 \right )
 \right ) + o(1),
\end{equation}
 when $z \rightarrow +\infty$.
 Like before,
 Eq.~\ref{eq: proof multivariate identification g>1 second eq} is incompatible with having $\alpha_g p_g^{(\gamma(t))} \neq \alpha_g' p_g^{'(\gamma(t))}$ for some $t$.
 If this were the case, Eq.~\ref{eq: proof multivariate identification g>1 second eq} would imply the following equation for the smallest such value of $t$, which is denoted by $t_0$.
%
%
\begin{equation*}
\resizebox{\textwidth}{!}
{$
\begin{split}
 \alpha_g \left ( 1+\sum_{t=t_0}^{|\Gamma|}
 p_g^{(\gamma(t))} \left ( e^{(|\Gamma|-t+1) z} -1 \right )
 \right ) e^{-(|\Gamma|-t_0+1) z} + o(1)  \quad =& \quad
 \alpha_g' \left ( 1+\sum_{t=t_0}^{|\Gamma|}
 p_g^{'(\gamma(t))} \left ( e^{(|\Gamma|-t+1) z} -1
 \right ) \right ) \times \\
 & \qquad e^{-(|\Gamma|-t_0+1) z} + o(1),
\end{split}
$}
\end{equation*}
 where the left-hand side goes to $\alpha_g p_g^{(\gamma(t_0))}$ and  the right-hand side goes to $\alpha_g' p_g^{'(\gamma(t_0))}$, and the two limits must coincide, when $z \rightarrow +\infty$.
 Since this contradicts the assumption about $t_0$, we must have $\alpha_g p_g^{(\gamma)} = \alpha_g' p_g^{'(\gamma)}$ for each $\gamma$, which implies that $\alpha_g=\alpha_g'$ (based on Eq.~\ref{eq: proof multivariate identification g>1 second eq}) and $p_g^{(\gamma)} = p_g^{'(\gamma)}$ for each $\gamma$.

\hfill QED

\vspace{6pt}

The next theorem extends Theorem 3 from \citeA{dasylva_goussanou_jjsds_2022}, which is about the consistency of the maximum likelihood estimators.
 In order to state this extension, more notation is needed.
 For $G \geq 1$, consider the finite mixture of $G$ distributions from ${\cal F}$, where $g$-th component has probability $\alpha_g$ and parameters $\bm{p}_g$ and $\bm{\lambda}_g$, and denote by
 $\bm{\theta} = \left [
  \left ( \alpha_g, \bm{p}_g, \bm{\lambda}_g \right )
  \right ]_{1 \leq g \leq G}$ the associated mixture parameters and by $q(.;\bm{\theta})$ the corresponding PMF.
%
%
\begin{equation}
\resizebox{\textwidth}{!}
{$
\begin{split}
 q \left ( \bm{n}; \bm{\theta} \right )
 \quad =& \quad
 \sum_{g=1}^G \alpha_g \left (
 I \left ( |\bm{n}|=0 \right )
 \left ( 1- \left | \bm{p}_g \right |\right )
 e^{- \left | \bm{\lambda}_g \right |} +
 I \left ( |\bm{n}|>1 \right )
 \left ( \left ( 1- \left | \bm{p}_g \right | \right )
 \prod_{\gamma \in \Gamma}
 \frac{e^{-\lambda_g^{(\gamma)}}
 \left ( \lambda_g^{(\gamma)}\right )^{n^{(\gamma)}}
 }{n^{(\gamma)}!} + \right . \right . \\
 & \qquad
 \left . \left . 
 \sum_{\gamma \in \Gamma: \ n^{(\gamma)}>0} p_g^{(\gamma)}
 \frac{e^{-\lambda_g^{(\gamma)}}
 \left ( \lambda_g^{(\gamma)}\right )^{n^{(\gamma)}-1}
 }{\left ( n^{(\gamma)} -1\right )!}
 \prod_{\gamma' \in \Gamma \setminus \{\gamma \}}
 \frac{e^{-\lambda_g^{(\gamma')}}
 \left ( \lambda_g^{(\gamma')}\right )^{n^{(\gamma')}}
 }{n^{(\gamma')}!} \right ) \right ), \
 \bm{n} = \left [ n^{(\gamma)}\right ]_{\gamma \in \Gamma} \in
 \mathbb{N}^{|\Gamma|},
\end{split}
$}
\end{equation}
 Define
%
%
\begin{equation}
 M_N (\bm{\theta}) =
 \frac{1}{|S_B|} \sum_{i \in S_B}
 \log q \left ( \bm{n}_i; \bm{\theta} \right ),
\end{equation}
 and for an integer $\tau_N>0$, define
%
%
\begin{equation}
\begin{split}
 M_N (\bm{\theta}; \tau_N) \quad =& \quad
 \frac{1}{|S_B|} \sum_{i \in S_B} \left (
 I \left ( |\bm{n}_i| \leq \tau_N \right ) 
 \log q \left ( \bm{n}_i; \bm{\theta} \right ) + \right . \\
 & \qquad
 \left . I \left ( |\bm{n}_i| \geq \tau_N+1 \right )
 \log \left ( \sum_{\bm{n} \in \mathbb{N}^{|\Gamma|}:
 \ |\bm{n}| \geq \tau_N+1}
 q \left ( \bm{n}; \bm{\theta} \right ) \right ) \right ).
\end{split}
\end{equation}

\noindent For
 $\bm{\theta}_{\varrho} =
  \left [ \left ( \alpha_g, \bm{\beta}_g, \bm{\lambda}_g \right )
  \right ]_{1 \leq g \leq G}$, let
 $\bm{\theta} \left ( \bm{\theta}_{\varrho} \right ) =
  \left [ \left ( \alpha_g, \varrho \left ( \bm{\beta}_g \right ),
  \bm{\lambda}_g \right ) \right ]_{1 \leq g \leq G}$.
 Also for a positive integer $d$, let $\bm{0}_d$ denote the $d$-tuple with all zeros.

As before, the mixture parameters may be estimated by maximizing the log-likelihood of the $\bm{n}_i$'s, i.e., $M_N (.)$ or $M_N (;\tau_N)$.
The following theorem states that the resulting estimators are consistent under suitable conditions, which include Eq.~\ref{eq: multivariate census regularity condition 2} or Eq.~\ref{eq: multivariate census regularity condition 2 parametric p}.
 In the latter case, it is assumed that the mapping $\varrho$ is injective.
%
%
\begin{theorem}\label{theorem: multivariate consistent estimation}
\noindent For $G^*\geq 2$ and $\nu \in (0,1)$, let $\Theta_1,\ldots,\Theta_{G^*}$ denote compact subsets of
 $\mathbb{R}^{(2|\Gamma|+1)G^{\ast}-1}$ such that
%
%
\begin{equation*}
\begin{split}
 \Theta_G \quad \subset& \quad
 \Bigg \{ \left [ \left ( \alpha_g,\bm{p}_g,\bm{\lambda}_g\right )
 \right ]_{1 \leq g \leq G^*} \in
 \mathbb{R}^{(2|\Gamma|+1)G^{\ast}-1} \ s.t. \\
 & \qquad \
 \left ( \alpha_g, \bm{p}_g, \bm{\lambda}_g \right ) \in
 (\nu,1]\times [0,1]^{|\Gamma|} \times [\nu,\Lambda]^{|\Gamma|}
 \ and \ \left | \bm{p}_g\right |\leq 1-\nu \ and \\
 & \qquad \
 \bm{\lambda}_{g+1} \succ \bm{\lambda}_{g}
 \ if \ g \geq G^*-G+1, \\
 & \qquad \
 \left ( \alpha_g, \bm{p}_g, \bm{\lambda}_g \right ) =
 (0,\bm{0}_{|\Gamma|},\bm{0}_{|\Gamma|})
 \ if \ g \leq G^*-G, \\
 & \qquad \
 \alpha_1+\ldots+\alpha_{G^{\ast}} = 1 \Bigg \},
 \ G=1,\ldots,G^{\ast}.
\end{split}
\end{equation*}
 and let $\Theta=\bigcup_{G=1}^{G^*} \Theta_G$.
 For an injective mapping $\varrho: \mathbb{R}^m \rightarrow [0,1]^{|\Gamma|}$, also let $\Theta_{\varrho 1},\ldots,\Theta_{\varrho G^*}$ denote compact subsets of $\mathbb{R}^{(|\Gamma|+m+1)G^{\ast}-1}$ such that
%
%
\begin{equation*}
\begin{split}
 \Theta_{\varrho G} \quad \subset& \quad
 \Bigg \{ \left [ \left ( \alpha_g,\bm{\beta}_g,\bm{\lambda}_g
 \right )\right ]_{1 \leq g \leq G^*} \in
 \mathbb{R}^{(|\Gamma|+m+1)G^{\ast}-1} \ s.t. \\
 & \qquad \
 \left ( \alpha_g, \bm{\beta}_g, \bm{\lambda}_g \right ) \in
 (\nu,1]\times \mathbb{R}^m \times [\nu,\Lambda]^{|\Gamma|}
 \ and
 \ \left | \varrho \left ( \bm{\beta}_g \right )\right |\leq 1-\nu
 \ and \\
 & \qquad \
 \bm{\lambda}_{g+1} \succ \bm{\lambda}_{g}
 \ if \ g \geq G^*-G+1, \\
 & \qquad \
 \left ( \alpha_g, \bm{\beta}_g, \bm{\lambda}_g \right ) =
 (0,\bm{0}_m,\bm{0}_{|\Gamma|})
 \ if \ g \leq G^*-G, \\
 & \qquad \
 \alpha_1+\ldots+\alpha_{G^{\ast}} = 1 \Bigg \},
 \ G=1,\ldots,G^{\ast},
\end{split}
\end{equation*}
 and let $\Theta_{\varrho}=\bigcup_{G=1}^{G^*} \Theta_{\varrho G}$.
 Suppose that all the linkage rules are simple (i.e., each rule is such that the decision to link two records involves no other record), C.1-C.3
 (from Theorem~\ref{theorem: law of large number}) apply, and Eq.~\ref{eq: multivariate census regularity condition 2} also applies with
%
%
\begin{equation*}
 F(\bm{p},\bm{\lambda}) =
 \sum_{g=1}^{G^{\ast}} \alpha_{0g} I \left (
 (\bm{p},\bm{\lambda})=
 \left ( \bm{p}_{0g},\bm{\lambda}_{0g}\right )\right ),
\end{equation*}
 $\bm{\theta}_0=
  \left [ \left ( \alpha_{0g},\bm{p}_{0g},\bm{\lambda}_{0g}\right )
  \right ]_{1\leq g \leq G^{\ast}} \in \Theta_{G_0}$ and
 $1 \leq G_0 \leq G^*$, and let
 $\widehat{\bm{\theta}}_{1N}$, $\widehat{\bm{\theta}}_{2N}$ and $\widehat{\bm{\theta}}_{3N}$ denote the estimators, which respectively maximize $M_N(.)$ over $\Theta_{G_0}$, $M_N(.)$ over $\Theta$, and $M_N(.;\tau_N)$ over $\Theta$, where $\tau_N$ is a positive integer such that $\tau_N \rightarrow \infty$ and $\tau_N = O \left ( \log N\right )$.
Then $\widehat{\bm{\theta}}_{1N}$, $\widehat{\bm{\theta}}_{2N}$ and $\widehat{\bm{\theta}}_{3N}$ converge in probability to $\bm{\theta}_0$.
 Furthermore, suppose that $\bm{p}_N(v)$ is also of the form $\varrho(\bm{\beta}_N(v))$ for $\bm{\beta}_N: {\cal V}_N^{\ast} \rightarrow \mathbb{R}^m$, which satisfies Eq.~\ref{eq: multivariate census regularity condition 2 parametric p} with
%
%
\begin{equation*}
 H(\bm{\beta},\bm{\lambda}) =
 \sum_{g=1}^{G^{\ast}} \alpha_{0g} I \left (
 (\bm{\beta},\bm{\lambda})=
 \left ( \bm{\beta}_{0g},\bm{\lambda}_{0g}\right )\right ),
\end{equation*}
 $\varrho \left ( \bm{\beta}_{0g}\right )=\bm{p}_{0g}$ for
 $g=G^{\ast}-G_0+1,\ldots,G^{\ast}$ and
 $\bm{\theta}_{\varrho 0}=
  \left [ \left ( \alpha_{0g},\bm{\beta}_{0g},
  \bm{\lambda}_{0g}\right ) \right ]_{1\leq g \leq G^{\ast}} \in
  \Theta_{\varrho G_0}$.
 Then, $\bm{\theta}_{\varrho 0}$ is also estimated consistently by maximizing $M_N(.)$ over $\Theta_{\varrho G_0}$, $M_N(.)$ over $\Theta_{\varrho}$, or $M_N(.;\tau_N)$ over $\Theta_{\varrho}$.
\end{theorem}


\noindent \underline{Proof}:
 The proof of the convergence of $\widehat{\bm{\theta}}_{1N}$, $\widehat{\bm{\theta}}_{2N}$ and $\widehat{\bm{\theta}}_{3N}$, adapts that of Theorem 3 in \citeA{dasylva_goussanou_jjsds_2022}, which consists in showing that $\bm{\theta}_0$ is an isolated maximum of the Kullback-Leibler divergence and that $M_N(.)$ and $M_N(.;\tau_N)$ converge uniformly over compact subsets containing $\bm{\theta}_0$.
 The adaptation involves a few changes, after which the previous proof carries over to the current setting.
 These changes include replacing each mention of Theorem 1, Lemma 1 and Lemma 4 in the previous proof by a mention of
 Theorem~\ref{theorem: law of large number},
 Lemma~\ref{lemma: multivariate convergence in distribution} and
 Lemma~\ref{lemma: identification multivariate}, respectively.
 They also involve the redefinition of $M(.)$, $R(.)$, $Z_i$ and $Z_i'$ as follows.
 For $M(.)$, let
%
%
\begin{equation*}
 M(\bm{\theta}) =
  \sum_{\bm{n} \in \mathbb{N}^{|\Gamma|}} q(\bm{n};\bm{\theta}_0)
  \log q(\bm{n};\bm{\theta}), \ \bm{\theta} \in \Theta.
\end{equation*}
 For $R(.)$, observe that
 $q \left ( \bm{n}; \bm{\theta} \right ) \leq 1$,
%
%
\begin{eqnarray*}
 q \left ( \bm{n}; \bm{\theta} \right ) &\geq&
 \alpha_{G^{\ast}-G+1}
 \left ( 1- \left | \bm{p}_{G^{\ast}-G+1} \right |\right )
 e^{- \left | \bm{\lambda}_{G^{\ast}-G+1} \right |} \times \\
 & &
 \left (  I \left ( |\bm{n}|=0 \right ) +
 I \left ( |\bm{n}|>1 \right )
 \prod_{\gamma \in \Gamma}
 \frac{ \left (
 \lambda_{G^{\ast}-G1}^{(\gamma)}\right )^{n^{(\gamma)}}}{n^{(\gamma)}!} \right ) \\
%
%
 &\geq&
 \nu^2 e^{-\Lambda}
 \left (  I \left ( |\bm{n}|=0 \right ) +
 I \left ( |\bm{n}|>1 \right )
 \prod_{\gamma \in \Gamma}
 \frac{ \left (
 \lambda_{G^{\ast}-G1}^{(\gamma)}\right )^{n^{(\gamma)}}}{n^{(\gamma)}!} \right ),
\end{eqnarray*}
 where $\bm{\theta} \in \Theta$.
 Hence
%
%
\begin{equation*}
 \log q \left ( \bm{n}; \bm{\theta} \right ) \geq
 \log \left ( \nu^2 e^{-\Lambda} \right ) +
 I \left ( |\bm{n}|>1 \right )
 \log \left ( \prod_{\gamma \in \Gamma}
 \frac{ \left (
 \lambda_{G^{\ast}-G1}^{(\gamma)}\right )^{n^{(\gamma)}}}{n^{(\gamma)}!}  \right ).
\end{equation*}
 Since the left-hand side is not positive, the above inequality implies that
%
%
\begin{eqnarray*}
 \left | \log q \left ( \bm{n}; \bm{\theta} \right ) \right | &\leq&
 \left | \log \left ( \nu^2 e^{-\Lambda} \right ) +
 I \left ( |\bm{n}|>1 \right )
 \log \left ( \prod_{\gamma \in \Gamma}
 \frac{ \left (
 \lambda_{G^{\ast}-G1}^{(\gamma)}\right )^{n^{(\gamma)}}}{n^{(\gamma)}!}  \right ) \right | \\
%
%
 &\leq&
 \left | \log \left ( \nu^2 e^{-\Lambda} \right ) \right | +
 \left | \log \left ( \prod_{\gamma \in \Gamma}
 \frac{ \left (
 \lambda_{G^{\ast}-G1}^{(\gamma)}\right )^{n^{(\gamma)}}}{n^{(\gamma)}!}  \right ) \right | \\
%
%
 &\leq&
 \left | \log \left ( \nu^2 e^{-\Lambda} \right ) \right | +
 \left | \log \left ( \prod_{\gamma \in \Gamma}
 \frac{ \left (
 \lambda_{G^{\ast}-G1}^{(\gamma)}\right )^{n^{(\gamma)}}}{n^{(\gamma)}!}  \right ) \right | \\
\end{eqnarray*}
 To find an upperbound on the above right-hand side, note that
 $\lambda_{G^{\ast}-G1}^{(\gamma)} \in [\nu,\Lambda]$. Thus
 \[ \left | \log \lambda_{G^{\ast}-G1}^{(\gamma)}  \right | \leq
    \max \left ( |\log \nu|, |\log \Lambda|\right ), \]
 which easily leads to
%
%
\begin{equation*}
\left | \log q \left ( \bm{n}_i; \bm{\theta} \right )  \right |
\leq
\left | \log \left ( \nu^2 e^{-|\Gamma| \Lambda}
\right ) \right | +
\max \left ( |\log \nu|,|\log \Lambda|\right )
\left | \bm{n}_i \right | + \left | \bm{n}_i \right |^2.
\end{equation*}
 Consequently, redefine $R(.)$ according to the right-hand side of the above equation, i.e.,
%
%
\begin{equation*}
R(t) =
2 \left | \log \left ( \nu^2 e^{-|\Gamma| \Lambda}
\right ) \right | +
\max \left ( |\log \nu|,|\log \Lambda|\right ) t + t^2,
 \ t=0,1,\ldots,
\end{equation*}
 and replace each mention of $R(n_i)$ in the previous proof by $R(|\bm{n}_i|)$.
 Finally, let
 $Z_i = \left | \log q \left ( \bm{n}_i; \bm{\theta} \right )
  \right |$ and
 $Z_i' = I \left ( |\bm{n}_i| \geq \tau_N+1\right )
  \left ( R(|\bm{n}_i|)+R(\tau_N+1)\right )$.

\noindent The proof of the consistent estimation of
 ${\bm{\theta}}_{\varrho 0}$ is similar based on
%
%
\begin{equation*}
 M(\bm{\theta}_{\varrho}) =
  \sum_{\bm{n} \in \mathbb{N}^{|\Gamma|}}
  q(\bm{n}; \bm{\theta}(\bm{\theta}_{\varrho 0}))
  \log q(\bm{n}; \bm{\theta}(\bm{\theta}_{\varrho})),
 \ \bm{\theta}_{\varrho} \in \Theta_{\varrho},
\end{equation*}
 $Z_i =
  \left | \log q \left ( \bm{n}_i; \bm{\theta}(\bm{\theta}_{\varrho})
  \right ) \right |$, with $R(.)$ and $Z_i'$ defined as above.

\hfill QED

\vspace{6pt}

In the above theorem, the mapping $\varrho(.)$ must be injective. The next lemma shows that this condition is met when $\varrho(.)$ is based on a nonsaturated log-linear specification of the interactions in the matched pairs.


\begin{lemma}\label{lemma: injective log-linear}

For positive integers $K,H_1,\ldots,H_K$, let
 $\Gamma =
  \{0,\ldots,H_1\}\times \ldots \times \{0,\ldots,H_K\}-\bm{0}_K$ and  $\bm{p} = \left [ p^{(\gamma)}\right ]_{\gamma \in \Gamma}$ be of the form $p^{(\gamma)} = P(i \in S_A) r^{(\gamma)}$, where
 $\sum_{\gamma \in \Gamma \cup \{ \bm{0}_K\}} r^{(\gamma)}=1$ and
 $r^{(\gamma)}$ has the following log-linear form with no interactions of order greater than $d<K$.
%
%
\begin{equation}
\label{eq: log-linear form}
 r^{(\gamma)} =
 \exp \left ( u + \sum_{t=1}^d
 \sum_{1 \leq k_1<\ldots<k_t \leq K}
 u_{k_1 \ldots k_t \left ( \gamma_{k_1}\ldots \gamma_{k_t} \right )}
 \right ),
\end{equation}
 where the term $u_{k_1 \ldots k_t \left ( \gamma_{k_1}\ldots \gamma_{k_t} \right )}$ is set to zero if one of  $\gamma_{k_1},\ldots ,\gamma_{k_t}$ is null, according to the dummy coding convention.
 Let $\bm{\beta}$ denote the vector that comprises $P(i \in S_A)$ and the parameters of $r^{(\gamma)}$, which are not set to zero by this convention.
 Then the mapping $\varrho: \bm{\beta} \mapsto \bm{p}$ is injective.

\end{lemma}


\noindent \underline{Proof}:
 We only need to show that $r^{(\gamma)}$ is uniquely determined by $\bm{p}$ for each $\gamma$.
 From the convention regarding $u_{k_1 \ldots k_t \left ( \gamma_{k_1}\ldots \gamma_{k_t} \right )}$, and the fact that $r^{(\gamma)}$ adds up to 1.0 over $\Gamma \cup \{\bm{0}_K\}$, it follows that $r^{(\gamma)}$ is a function of
 $\sum_{t=1}^d
  \sum_{1 \leq k_1<\ldots<k_t \leq K} H_{k_1}\ldots H_{k_t}$
 free parameters given by
%
%
\begin{equation*}
 \left [ \left (
  u_{k_1 \ldots k_t \left ( \gamma_{k_1}\ldots \gamma_{k_t} \right )}
  \right )_{
  \begin{array}{l}
  \scriptstyle 1 \leq k_1<\ldots<k_t \leq K,\\
  \scriptstyle
  \left ( \gamma_{k_1}\ldots \gamma_{k_t} \right )
  \in \{1,\ldots,H_{k_1}\}\times \ldots \times \{1,\ldots,H_{k_t}\}
  \end{array}}
  \right ]_{t=1,\ldots,d},
\end{equation*}
 with
%
%
\begin{eqnarray*}
 \log r (\gamma) &=&
 u + \sum_{t=1}^d \sum_{1 \leq k_1<\ldots<k_t \leq K}
 I \left ( \gamma_{k_1},\ldots,\gamma_{k_t}>0 \right )
 u_{k_1 \ldots k_t
 \left ( \gamma_{k_1}\ldots \gamma_{k_t} \right )}, \\
 e^{-u} &=&
 \left ( 1 + \sum_{\gamma \in \Gamma}
 \exp \left ( \sum_{t=1}^d \sum_{1 \leq k_1<\ldots<k_t \leq K}
 I \left ( \gamma_{k_1},\ldots,\gamma_{k_t}>0 \right )
 u_{k_1 \ldots k_t
 \left ( \gamma_{k_1}\ldots \gamma_{k_t} \right )} \right ) 
 \right )^{-1}.
\end{eqnarray*}
 For integers $1 \leq \kappa_1 < \ldots < \kappa_{d+1} \leq K$, and
 $\widetilde{\gamma}_{\kappa_1},\ldots,
  \widetilde{\gamma}_{\kappa_{d+1}}$,
 such that
 $1 \leq \widetilde{\gamma}_{ \kappa_t } \leq
  H_{\kappa_t }$, for $t=1,\ldots,d+1$, let
%
%
\begin{equation*}
 \widetilde{\Gamma} =
 \left \{ \gamma \in \Gamma \ s.t. \
 \gamma_k=0 \ if \ k \notin \{\kappa_1,\ldots,\kappa_{d+1}\},
 \ and \ \gamma_{\kappa_t} \in \{0, \widetilde{\gamma}_{\kappa_t} \}
 \ for \ t = 1,\ldots,d+1
 \right \},
\end{equation*}
 and let $\widetilde{\gamma}_k=0$ if
 $k \notin \{\kappa_1,\ldots,\kappa_{d+1}\}$, and $\widetilde{\gamma}_k=\widetilde{\gamma}_{\kappa_t}$ if
 $k=\kappa_t$ for some $t$.

\noindent To show that the free parameters are uniquely determined by
 $\bm{p}$, it is sufficient to show that the parameters
%
%
\begin{equation*}
 \left [ \left (
  u_{k_1 \ldots k_t
  \left ( \widetilde{\gamma}_{k_1}\ldots
  \widetilde{\gamma}_{k_t} \right )}
  \right )_{\{k_1,\ldots,k_t\} \subset
  \{\kappa_1,\ldots,\kappa_{d+1}\}: \ k_1<\ldots<k_t}
 \right ]_{t=1,\ldots,d}
\end{equation*}
 are uniquely determined by
 $\left [ {p}^{(\gamma)}
  \right ]_{\gamma \in \widetilde{\Gamma}}$ for each choice of $\{\kappa_1,\ldots,\kappa_{d+1}\}$ and $\widetilde{\gamma}_{\kappa_1},\ldots,\widetilde{\gamma}_{d+1}$.
 For such a choice, let $\gamma^{\ast}$ denote the element of $\widetilde{\Gamma}$ such that $\gamma^{\ast}_{\kappa_t}=\widetilde{\gamma}_{\kappa_t}$ for $t=1,\dots,d+1$, and let $q^{(\gamma)} = {p}^{(\gamma)}/{p}^{(\gamma^{\ast})}$ for $\gamma \in \widetilde{\Gamma}$.
 Also, let
%
%
\begin{equation}
 \label{eq: proof log-linear injective}
 s =
 \sum_{t=1}^d
 \sum_{ \{k_1,\ldots,k_t\} \subset
  \{\kappa_1,\ldots,\kappa_{d+1}\}: \ k_1<\ldots<k_t}
  u_{k_1 \ldots k_t
  \left ( \widetilde{\gamma}_{k_1}\ldots
  \widetilde{\gamma}_{k_t} \right )}.
\end{equation}
 It is next shown by induction on $t=1,\ldots,d,$ that there exists a function $a_t(.;.)$ such that
%
%
\begin{equation}
\label{eq: induction property}
 u_{k_1 \ldots k_t
  \left ( \widetilde{\gamma}_{k_1}\ldots
  \widetilde{\gamma}_{k_t} \right )} =
 (-1)^{t+1} s +
 \sum_{\gamma \in \widetilde{\Gamma}:
 \ \sum_{k=1}^K I(\gamma_k>0)\leq t}
 a_t \left ( \gamma; k_1, \ldots, k_t \right )\log q(\gamma),
\end{equation}
 for each subset
 $\{k_1,\ldots,k_t\}$ of
 $\{\kappa_1,\ldots,\kappa_{d+1}\}$ such that $k_1<\ldots<k_t$.

\noindent Indeed, for $t=1$ and $\gamma \in \widetilde{\Gamma}$ such that $\sum_{k=1}^K I(\gamma_k>0)=1$, we have
%
%
\begin{equation*}
 \log q(\gamma) =
 \sum_{k \in \{\kappa_1,\ldots,\kappa_{d+1}\}}
 I(\gamma_k>0) u_{k(\widetilde{\gamma}_k)} - s.
\end{equation*}
 Therefore
%
%
\begin{equation*}
 u_{k(\widetilde{\gamma}_k)} =
 s  +
 \sum_{\gamma \in \widetilde{\Gamma}:
 \ \sum_{k=1}^K I(\gamma_k>0) \leq 1}
 I(\gamma_k=\widetilde{\gamma}_k)\log q(\gamma),
 \ k \in \{\kappa_1,\ldots,\kappa_{d+1}\},
\end{equation*}
 and the property is true with
 $a_1(\gamma;k)=I(\gamma_k=\widetilde{\gamma}_k)$.

\noindent Now, suppose that the property holds up to $t<d$.
 Then for a subset $\{k_1,\ldots,k_{t+1}\}$ of
 $\{\kappa_1,\ldots,\kappa_{d+1}\}$ such that $k_1<\ldots<k_{t+1}$, and $\gamma \in \widetilde{\Gamma}$ such that
 $\sum_{k=1}^K I(\gamma_k>0)=t+1$, we have
%
%
\begin{eqnarray*}
 \log q (\gamma) &=&
 -s +
 \sum_{t'=1}^t
 \sum_{\{k_1',\ldots,k_{t'}'\} \subset
  \{k_1,\ldots,k_{t+1}\}: \ k_1'<\ldots<k_{t'}'}
 u_{k_1' \ldots k_{t'}'
  ( \widetilde{\gamma}_{k_1'}\ldots
  \widetilde{\gamma}_{k_{t'}'} )} +
 u_{k_1 \ldots k_{t+1}
  ( \widetilde{\gamma}_{k_1}\ldots
  \widetilde{\gamma}_{k_{t+1}} )}.
\end{eqnarray*}
 Therefore
%
%
\begin{eqnarray*}
 u_{k_1 \ldots k_{t+1}
  ( \widetilde{\gamma}_{k_1}\ldots
  \widetilde{\gamma}_{k_{t+1}} )} &=&
 \underbrace{\left ( \sum_{t'=0}^t {t+1 \choose t'} (-1)^{t'}
 \right )}_{=-{t+1 \choose t+1}(-1)^{t+1} = (-1)^{(t+1)+1}} s + 
%
%
 \sum_{\gamma \in \widetilde{\Gamma}:
 \ \sum_{k=1}^K I(\gamma_k>0)\leq t} \left (
 -\sum_{t'=\sum_{k=1}^K I(\gamma_k>0)}^t \right . \\
 & &
 \underbrace{\left .
 \sum_{\{k_1',\ldots,k_{t'}'\} \subset
  \{k_1,\ldots,k_{t+1}\}: \ k_1'<\ldots<k_{t'}'}
 a_{t'} \left ( \gamma; k_1', \ldots, k_{t'}' \right )
 \right )}_{=a_{t+1} \left ( \gamma;k_1,\ldots,k_{t+1}\right )}
 \log q(\gamma) + \\
 & &
 \sum_{\gamma \in \widetilde{\Gamma}:
 \ \sum_{k=1}^K I(\gamma_k>0) = t+1}
 \underbrace{
 I \left ( \gamma_{k_1},\ldots,\gamma_{k_{t+1}} > 0\right )}_{
 =a_{t+1} \left ( \gamma;k_1,\ldots,k_{t+1}\right )}
 \log q(\gamma).
\end{eqnarray*}
 Consequently the property holds for $t+1$ and for each $t$ from $1$ to $d$.
 With Eq.~\ref{eq: proof log-linear injective}, this implies that
%
%
\begin{eqnarray*}
 s &=&
 (-1)^d \left ( \sum_{t=1}^d
 \sum_{ \{k_1,\ldots,k_t\} \subset
 \{\kappa_1,\ldots,\kappa_{d+1}\}: \ k_1<\ldots<k_t}
 \sum_{\gamma \in \widetilde{\Gamma}:
 \ \sum_{k=1}^K I(\gamma_k>0)\leq t}
 a_t \left ( \gamma; k_1, \ldots, k_t \right )\log q(\gamma) \right ).
\end{eqnarray*}
 Thus, $s$ is uniquely determined by $\left [ {p}^{(\gamma)}
  \right ]_{\gamma \in \widetilde{\Gamma}}$ and so are
%
%
\begin{equation*}
 \left [ \left (
  u_{k_1 \ldots k_t
  \left ( \widetilde{\gamma}_{k_1}\ldots
  \widetilde{\gamma}_{k_t} \right )}
  \right )_{\{k_1,\ldots,k_t\} \subset
  \{\kappa_1,\ldots,\kappa_{d+1}\}: \ k_1<\ldots<k_t}
 \right ]_{t=1,\ldots,d},
\end{equation*}
 according to Eq.~\ref{eq: induction property}.

\hfill QED


\section{Initialization procedure}\label{appendix: initialization}

This section describes the initialization procedure for fitting the multivariate neighbor model in the simulations.
 The model parameters include the mixing proportions (the $\alpha_g$'s), the parameters of the false positives distribution (the $\bm{\lambda}_g$'s) and those for the true positives distribution (the common value of the $\bm{p}_g$'s).
 With $G$ classes, the mixing proportion $\alpha_g$ is set to $1/G$ for each class.
 The other starting values are chosen as follows.

For the false positives distribution, each $\bm{\lambda}_g$ is set to a common value that is denoted by $\widehat{\bm{\lambda}} =
 \left [ \widehat{\lambda}^{(\gamma)}\right ]_{\gamma \in \Gamma}$, where $\Gamma = \{0,1\}^3 \setminus \{(0,0,0)\}$.
 For $\gamma \in \Gamma$, $\widehat{\lambda}^{(\gamma)}$ is set to the estimate of $\overline{\lambda}$ that is obtained by fitting the univariate  neighbor model, when the records are linked according to $\gamma$.
 For example, when $\gamma=(1,0,1)$, this means linking a pair if it is linked by the first linkage rule, and there is exact agreement on the surname and birth month but disagreement on the birth day.

 Based on $\widehat{\bm{\lambda}}$, the starting values are chosen for the coverage probability $P(i \in S_A)$ and the log-linear parameters (i.e., $u_{1(1)}$, $u_{2(1)}$, $u_{3(1)}$, $u_{12(11)}$, $u_{13(11)}$ and $u_{23(11)}$).
 These values are found in three steps as follows.
 In the first step, an estimate $\widehat{\bm{p}}=\left [ \widehat{p}^{(\gamma)}\right ]_{\gamma \in \Gamma}$ of the vector of true positives probabilities is computed by maximizing the log-likelihood of the multivariate model with a single class, where $\widehat{\bm{\lambda}}$ is plugged in, and the true positives probabilities are not constrained to have a log-linear form.
 This step corresponds to a convex optimization, because the log-likelihood of the multivariate model is concave with respect to the true positives probabilities, when the other parameters are given.
 In the second step, the starting values for the log-linear parameters of $r^{(\gamma)}$ (in Eq.~\ref{eq: postulated gamma distribution}) are found by a method of moments, based on $\widehat{\bm{p}}$ as follows.
 When fitting the model without interactions, let $\widehat{q}_k = \sum_{\gamma \in \Gamma: \ \gamma_k=1} \widehat{p}^{(\gamma)}$ (for $k=1,2,3$) and $\widehat{q}_{k_1 k_2}= \sum_{\gamma \in \Gamma: \ \gamma_{k_1}=\gamma_{k_2}=1} \widehat{p}^{(\gamma)}$ (for $1 \leq k_1 < k_2 \leq 3$) and choose the starting values as
\begin{eqnarray*}
 \widehat{u}_{1(1)} &=&
 logit \left ( \frac{1}{2} \left ( \frac{\widehat{q}_{12}}{\widehat{q}_2} +
 \frac{\widehat{q}_{13}}{\widehat{q}_3} \right ) \right ),\\
 \widehat{u}_{2(1)} &=&
 logit \left ( \frac{1}{2} \left ( \frac{\widehat{q}_{12}}{\widehat{q}_1} +
 \frac{\widehat{q}_{23}}{\widehat{q}_3} \right ) \right ),\\
 \widehat{u}_{3(1)} &=&
 logit \left ( \frac{1}{2} \left ( \frac{\widehat{q}_{13}}{\widehat{q}_1} +
 \frac{\widehat{q}_{23}}{\widehat{q}_2} \right ) \right ),
\end{eqnarray*}
 with $\widehat{u}_{12(11)}=\widehat{u}_{13(11)}=\widehat{u}_{23(11)}=0$.
 When including the interactions, choose the starting values as
%
%
\begin{equation*}
\resizebox{\textwidth}{!}
{$
\begin{split}
 \widehat{u}_{1(1)} \quad =& \quad
 \log \left (\frac{\widehat{p}^{(1,1,0)}}{
 \widehat{p}^{(1,1,1)} } \right ) +
 \log \left (\frac{\widehat{p}^{(1,0,1)}}{
 \widehat{p}^{(1,1,1)} } \right ) +
 \log \left (\frac{\widehat{p}^{(0,1,1)}}{
 \widehat{p}^{(1,1,1)} } \right ) -
 \left ( \log \left (\frac{\widehat{p}^{(1,0,0)}}{
 \widehat{p}^{(1,1,1)} } \right ) +
 \log \left (\frac{\widehat{p}^{(0,1,0)}}{
 \widehat{p}^{(1,1,1)} } \right ) \right ), \\
%
%
 \widehat{u}_{2(1)} \quad =& \quad
 \log \left (\frac{\widehat{p}^{(1,1,0)}}{
 \widehat{p}^{(1,1,1)} } \right ) +
 \log \left (\frac{\widehat{p}^{(1,0,1)}}{
 \widehat{p}^{(1,1,1)} } \right ) +
 \log \left (\frac{\widehat{p}^{(0,1,1)}}{
 \widehat{p}^{(1,1,1)} } \right ) -
 \left ( \log \left (\frac{\widehat{p}^{(1,0,0)}}{
 \widehat{p}^{(1,1,1)} } \right ) +
 \log \left (\frac{\widehat{p}^{(0,0,1)}}{
 \widehat{p}^{(1,1,1)} } \right ) \right ),\\
%
%
 \widehat{u}_{3(1)} \quad =& \quad
 \log \left (\frac{\widehat{p}^{(1,1,0)}}{
 \widehat{p}^{(1,1,1)} } \right ) +
 \log \left (\frac{\widehat{p}^{(1,0,1)}}{
 \widehat{p}^{(1,1,1)} } \right ) +
 \log \left (\frac{\widehat{p}^{(0,1,1)}}{
 \widehat{p}^{(1,1,1)} } \right ) -
 \left ( \log \left (\frac{\widehat{p}^{(0,1,0)}}{
 \widehat{p}^{(1,1,1)} } \right ) +
 \log \left (\frac{\widehat{p}^{(0,0,1)}}{
 \widehat{p}^{(1,1,1)} } \right ) \right ),\\
%
%
 \widehat{u}_{12(11)} \quad =& \quad
 -\left ( \log \left (\frac{\widehat{p}^{(1,1,0)}}{
 \widehat{p}^{(1,1,1)} } \right ) +
 \log \left (\frac{\widehat{p}^{(1,0,1)}}{
 \widehat{p}^{(1,1,1)} } \right ) +
 \log \left (\frac{\widehat{p}^{(0,1,1)}}{
 \widehat{p}^{(1,1,1)} } \right ) \right ) +
 \log \left (\frac{\widehat{p}^{(1,0,0)}}{
 \widehat{p}^{(1,1,1)} } \right ),\\
%
%
 \widehat{u}_{13(11)} \quad =& \quad
 -\left ( \log \left (\frac{\widehat{p}^{(1,1,0)}}{
 \widehat{p}^{(1,1,1)} } \right ) +
 \log \left (\frac{\widehat{p}^{(1,0,1)}}{
 \widehat{p}^{(1,1,1)} } \right ) +
 \log \left (\frac{\widehat{p}^{(0,1,1)}}{
 \widehat{p}^{(1,1,1)} } \right ) \right ) +
 \log \left (\frac{\widehat{p}^{(0,1,0)}}{
 \widehat{p}^{(1,1,1)} } \right ),\\
%
%
%
 \widehat{u}_{23(11)} \quad =& \quad
 -\left ( \log \left (\frac{\widehat{p}^{(1,1,0)}}{
 \widehat{p}^{(1,1,1)} } \right ) +
 \log \left (\frac{\widehat{p}^{(1,0,1)}}{
 \widehat{p}^{(1,1,1)} } \right ) +
 \log \left (\frac{\widehat{p}^{(0,1,1)}}{
 \widehat{p}^{(1,1,1)} } \right ) \right ) +
 \log \left (\frac{\widehat{p}^{(0,0,1)}}{
 \widehat{p}^{(1,1,1)} } \right ).
\end{split}
$}
\end{equation*}
 Finally,  let $\widehat{\bm{r}} = \left [ \widehat{r}^{(\gamma)} \right ]_{\gamma \in \{0,1\}^3}$ denote the vector of probabilities that correspond to the above starting values of the log-linear parameters (i.e., $\widehat{r}^{(\gamma)}$ is equal to the right-hand side of Eq.~\ref{eq: postulated gamma distribution}, where the starting values are plugged in) and choose the starting coverage as
\begin{equation*}
 \widehat{P}(i \in S_A) =
 \frac{\sum_{\gamma \in \Gamma} \widehat{p}^{(\gamma)}}{
 \sum_{\gamma \in \Gamma} \widehat{r}^{(\gamma)}}.
\end{equation*}

\end{appendix}

\end{document}